\documentclass[final, 5p, times, twocolumn]{elsarticle}

\usepackage[pdftex,dvipsnames]{xcolor}  
\usepackage{upgreek}
\usepackage{graphicx}         
\usepackage{amsmath}
\usepackage{amsfonts}
\usepackage[linesnumbered,ruled,vlined]{algorithm2e}
\usepackage{bm}   
\usepackage{url}
\usepackage{amssymb}
\usepackage{upgreek}
\usepackage{dsfont}
\usepackage{enumerate}

\newtheorem{proposition}{Proposition}
\newtheorem{theorem}{Theorem}
\newtheorem{lemma}{Lemma}
\newtheorem{problem}{Problem}
\newdefinition{defn}{Definition}
\newdefinition{rem}{Remark}
\newdefinition{assum}{Assumption}
\newproof{pf}{Proof}

\def\changes#1{\textcolor{black}{#1}}
\newcommand{\mfs}[1]{{\normalfont\textsf{#1}}}
\newcommand{\nd}{\text{d}}
\newcommand{\mf}{\mathbf}

\newcommand{\cov}{\normalfont\textsf{\footnotesize cov}}
\newcommand{\assign }{\leftarrow}

\newcommand{\tr}{\normalfont\textsf{\footnotesize tr}}
\newcommand{\vect}{\normalfont\textsf{\footnotesize vec}}

\newcommand{\diag}{\text{\normalfont  diag}}

\newcommand{\len}{\normalfont\textsf{\small len}}

\def\RevOne#1{\textcolor{Black}{#1}}
\def\RevTwo#1{\textcolor{Black}{#1}}
\def\RevFour#1{\textcolor{Black}{#1}}
\def\RevFive#1{\textcolor{Black}{#1}}
\def\RevAll#1{\textcolor{Black}{#1}}
\def\SecondRound#1{\textcolor{Black}{#1}}

\begin{document}

\begin{frontmatter}
\title{PLATE: a perception-latency aware estimator \tnoteref{t1}} 

\tnotetext[t1]{\textcolor{red}{This is the accepted version of the manuscript: R. Aldana-López, R. Aragüés, and C. Sagüés, “PLATE: A perception-latency aware estimator,” ISA Transactions, vol. 142, pp. 716-730, 2023, ISSN 0019-0578. doi: 10.1016/j.isatra.2023.08.013.
    \textbf{Please cite the publisher's version}. For the publisher's version and full citation details see:\\\protect\url{https://doi.org/10.1016/j.isatra.2023.08.013}.}}

\author[First]{Rodrigo Aldana-López*} 
\author[First]{Rosario Aragüés} 
\author[First]{Carlos Sagüés} 

\address[First]{Departamento de Informatica e Ingenieria de Sistemas (DIIS) and Instituto de Investigacion en Ingenieria de Aragon (I3A), 
\\
Universidad de Zaragoza, Zaragoza 50018, Spain.\\
(e-mail: rodrigo.aldana.lopez@gmail.com, raragues@unizar.es, csagues@unizar.es)}

\begin{abstract}       
Target tracking is a popular problem with many potential applications. There has been a lot of effort on improving the quality of the detection of targets using cameras through different techniques. In general, with higher computational effort applied, i.e., a longer perception-latency, a better detection accuracy is obtained. However, it is not always useful to apply the longest perception-latency allowed, particularly when the environment doesn't require to and when the computational resources are shared between other tasks. In this work, we propose a new Perception-LATency aware Estimator (PLATE), which uses different perception configurations in different moments of time in order to optimize a certain performance measure. This measure takes into account a perception-latency and accuracy trade-off aiming for a good compromise between quality and resource usage. Compared to other heuristic frame-skipping techniques, PLATE comes with a formal complexity and optimality analysis. \RevAll{The advantages of PLATE are verified by several experiments including an evaluation over a standard benchmark with real data and using state of the art deep learning object detection methods for the perception stage.}
\end{abstract}

\end{frontmatter}

\section{Introduction}

The Target Tracking (TT) problem involves using vision or range sensors to locate and track the target's trajectory. Sequential or online tracking aims to provide real-time estimates of the target's state as measurements are obtained \cite{luo2021}. This problem has a variety of applications, including automated surveillance \cite{Hu2004}, crowd analysis and pedestrian intention prediction \cite{madrigal2017}, gesture recognition \cite{mitra2007}, and traffic monitoring \cite{sokemi2016}. TT is also important in the robotics community, where it is used in robot trajectory tracking \cite{chen2017a}, cooperative surveillance \cite{cyber_cooperative}, and object tracking for localization and mapping \cite{bescos2021}.
Most TT solutions rely on a detector, which can be based on traditional vision methods \cite{yilmaz2006} or Deep Learning (DL) techniques \cite{redmon2016,beery2020}. When multiple targets are involved, an association and re-identification stage is added to distinguish detections between targets. Furthermore, probabilistic inference techniques such as Kalman filtering, particle filtering, or other Bayesian methods are typically used for model-based tracking \cite{luo2021}.

In this work, we adopt the term perception-latency from \cite{falanga2019} to refer to the time elapsed from the instant the sensors start sampling to the moment the target is detected and identified from the raw sensor data. Note that the perception-latency can be varied by changing the perception configuration, which includes the resolution of visual or range sensors, the number of features considered in the perception algorithm, and model hyperparameters if DL detectors are used. In many TT approaches, the perception configuration can be modified to improve accuracy at the expense of a longer perception-latency \cite{huang2017}, leading to a perception-latency/accuracy trade-off.  \RevTwo{For radar-based perception, the term dwell-time refers to the time taken by the radar to acquire a measurement \cite{narykov2013}, which is considered part of the perception latency and introduces a similar trade-off between quality and duration.}

\RevOne{Other works, such as \cite{gatsis2021}, have studied a similar latency/accuracy trade-off in networked control systems where the latency arises from the communication channel. While there are some similarities between this problem and the present work, our interest lies in the setting where the sensor, perception algorithm, and estimation algorithm are all in the same computing unit. In this case, the latency due to information transmission is negligible when compared to the perception latency, which is the primary concern.}

In addition, real-time tracking is often required, meaning that the perception-latency is subject to a real-time constraint determined by the sampling period of the vision sensors, typically $1/30s$ or $1/60s$ for traditional vision cameras \cite{fan2017}. Real-time TT is particularly relevant for resource-constrained platforms such as mobile systems \cite{howard2017} and smart cameras \cite{casares2011}. The perception-latency/accuracy trade-off is typically addressed by selecting the optimal perception configuration that satisfies the real-time constraint. Most approaches using DL involve testing different neural network architectures to find a suitable accuracy compromise with smaller models and less perception-latency.

Moreover, note that perception is not always the only task running in the system. Robotic platforms use computing resources for motion planning, control and coordination tasks. Hence, setting the real-time constraint for the perception task as the full sampling period of the vision sensor is still unrealistic. In addition, as pointed out in \cite{guan2018}, it is not always necessary to run the perception task for each new frame if the dynamics of the scene doesn't require to. Instead, one may use a frame-skipping policy obtaining a better resource usage. This means that CPU load for vision-based perception  must be managed as well.

In this sense, looking for adaptive perception schemes, in which perception-latency changes according to the situation, might be beneficial for a better resource usage under the perception-latency/accuracy trade-off. Some DL techniques allow adaptive forward latency such as the so called Anytime Neural Networks (ANN). ANNs provide a way to select online between different compromises of latency and accuracy \cite{hu2019}. These type of networks were used in \cite{yao2020} in order to schedule different DL jobs with latency chosen such that real-time deadlines are maintained with the best possible accuracy overall.  However, being a general DL scheduler, \cite{yao2020} does not take into account that for a tracking problem, while a longer latency might lead to a better accuracy for the detector, the model uncertainty might increase as well due to the delay introduced by the perception-latency \cite{aldana2020}.

Instead of using an adaptive perception-latency, some works have used frame-skipping techniques in order to reduce computational load. In \cite{guan2018}, the authors propose an event-triggered rule in which the detection is only updated under certain events, based on the expected motion of the target. This approach reduces the computational burden of the perception task by effectively skipping frames when the expected motion of the target doesn't require to. Other approaches follow a similar idea such as the one in \cite{casares2011} in which the perception task goes to an idle state depending on the expected state of the environment, obtaining an energy efficient use of battery powered smart cameras. Similarly, in \cite{luo2019} a neural network is trained based on examples in order to make a decision on whether or not the detection for some target should be updated. Although some of these methods shows good experimental performance in certain scenarios, they are mostly based in several heuristic rules from which it is hard to provide formal guarantees on the impact over the resource usage and quality for TT in the general case. 

\changes{In addition, the related problem of multi-sensor scheduling problem can be relevant in the perception scheduling context \cite{isa3,isa1}. In this problem, multiple sensors are available to measure the state of a dynamical system and the goal is to schedule them for control or estimation purposes under a given cost function. As discussed in \cite{ss_periodic_ih} this problem is usually subject to a combinatorial explosion in terms of the number of available sensors and the time horizon for the cost function. Hence, most efforts in the literature have focused on obtaining efficient sub-optimal solutions. For instance, \cite{ss_pruning} take advantage of the structure of the cost function and the system to prune many of the possible schedules. Despite this, while this approach reduces the complexity, the problem may still be intractable for big horizon sizes. Other approaches adopt a greedy policy which picks the best sensor at each time which optimizes a one-step look-ahead cost as in \cite{ss_greedy}. Nonetheless, despite the simplicity of this approach, the result tends to be quite conservative since the behaviour on the long run is not taken into account. }

\changes{On the other hand, another approach is to avoid the combinatorial explosion by restricting only to periodic schedules as in \cite{ss_periodic_filter,ss_periodic_ih,isa2}. However, this is conservative as well, being sub-optimal when not in steady state. Moreover, in most cases the actual optimal period for the schedule is unknown and depends on the particular problem setting. Other efficient approaches where proposed in \cite{ss_fast,ss_fast2} which do not require the periodic schedule assumption. Nonetheless, the actual form of the scheduling policies in the previous works is tightly coupled to the particular structure of the systems under consideration and the cost function. In this sense, it is not possible to apply directly these ideas when we take into account the perception-latency trade-off as well as computational load and energy.}

\RevAll{Motivated by this, our contribution is a new Perception-LATency aware Estimator (PLATE). PLATE schedules different perception configurations in different moments of time in order to optimize a performance measure that takes into account the perception-latency/accuracy trade-off aiming to improve estimation quality and resource usage. We show that the proposed perception-latency scheduling policy is not subject to a combinatorial explosion and approximates the optimal perception schedule with arbitrary precision as more computing time is employed. } \RevAll{The advantages of PLATE over the state of the art are summarized as follows
\begin{itemize}
\item Compared to the rest of the literature, PLATE explicitly takes into account the effect of the delay introduced by the perception-latency on the target uncertainty model. Moreover, the algorithm is not tightly coupled to the form of the cost function, making it more general and versatile than related approaches. This allows the use of PLATE in an online setting even when the perception quality is estimated to be different from its nominal values.
\item Compared to existing heuristic frame-skipping techniques \cite{guan2018,casares2011,luo2021}, we provide a theoretical analysis of our proposal, equipped with formal guarantees under linear model and additive Gaussian disturbance assumptions.
\item  Compared to multi-sensor scheduling approaches, we do not assume periodicity of the schedule, allowing to less conservative results, which can adapt to varying perception quality. 
\end{itemize}}
\RevAll{The advantages of PLATE are verified by several experiments including an evaluation over a standard benchmark with real data and using state of the art DL methods for the perception stage.}
\subsection{Notation}
Let $\bar{\mathbb{R}}_{\text{\fontsize{1}{0}$\geq \!\!0$}}=\mathbb{R}_{\text{\fontsize{1}{0}$\geq \!\!0$}}\cup \{\infty\}$. Moreover, let $\mathbb{E}\{\bullet\}$ be the expectation operator and $\cov\{\bullet,\ast\} = \mathbb{E}\{ (\bullet-\mathbb{E}\{\bullet\})(\ast-\mathbb{E}\{\ast\})^\top \}$ and $\cov\{\bullet\}:=\cov\{\bullet,\bullet\}$. Let $\tr(\bullet)$ and $\vect(\bullet)$ be the trace and vectorization operators respectively. Let $\otimes$ denote the Kronecker product. Let $\diag(v_1,\dots,v_n)\in\mathbb{R}^{n\times n}$ be a matrix with diagonal components $v_1,\dots,v_n\in\mathbb{R}$. We use $\mf{x}(t)$ to denote the evaluation of a signal $\mf{x}$ at continuous time $t$ whereas $\mf{x}[k]:=\mf{x}(\tau_k)$ for discrete time instants $\{\tau_k\}_{k=0}^\infty$. In addition, for a sequence $p=\{p_k\}_{k=0}^{\ell-1}$ of length $\ell\in\mathbb{N}\cup\{\infty\}$, denote $\len(p) := \ell$. For any matrix $\mf{A}\in\mathbb{R}^{n\times n}$ with components $[\mf{A}]_{ij}\in\mathbb{R}$, let $\|\mf{A}\|_F = \sqrt{\sum_{i=1}^n\sum_{j=1}^n [\mf{A}]_{ij}^2}$ its Frobenius norm \cite[Page 341]{horn}. Let $\mf{A}\succ \mf{B}$ or $\mf{A}\succeq \mf{B}$ denote positive definiteness and semi-definiteness of $\mf{A}-\mf{B}$ respectively. \RevFive{For a set $\bullet$, let $\min\{\bullet\}, \max\{\bullet\}, \inf(\bullet), \sup(\bullet)$ represent the minimum, maximum, infinimum and supremum standard set operators.}

\section{Problem statement}
\label{sec:problem}
Consider the following target model, with state $\mf{x}(t)\in\mathbb{R}^{n_\mf{x}}$ e.g. containing its position and velocity. Assume a simplified motion model given by the following Stochastic Differential Equation (SDE):
\begin{equation}
\label{eq:system_sde}
    \nd \mf{x}(t) = \mf{A}\mf{x}(t)\nd t + \mf{B}\nd \mf{w}(t), \ \ t\geq 0
\end{equation}
where $\mf{A}\in\mathbb{R}^{{n_\mf{x}}\times {n_\mf{x}}},\mf{B}\in\mathbb{R}^{{n_\mf{x}}\times {n_\mf{w}}}$ and $\mf{w}(t)$ is a ${n_\mf{w}}$-dimensional Wiener processes with covariance $\cov\{\mf{w}(s),\mf{w}(r)\} = \mf{W}\min(s,r)$ \cite[Page 63]{astrom}. As usual, the process $\mf{w}(t)$ models disturbances, unknown inputs for the target and non-modeled dynamics. Moreover, $\mf{x}(0)$ is normally distributed with mean $\mf{x}_0$ and covariance $\mf{P}_0$. 

The goal is to construct an estimation framework for the state $\mf{x}(t)$  using available sensors, e.g. vision or range. The sensors can produce \textit{raw measurements} with a minimum sampling period of $\Delta_s$. In order to use these measurements, they must be processed. For each raw measurement, the
system can choose between $D$ different perception methods
which take the raw measurements to produce a \textit{processed
measurement} for the position of the target through a detection
process. Thus, processed measurements are available to be used by the system only at, perhaps non-uniform, instants $\{\tau_k\}_{k=0}^\infty$ all integer multiples of $\Delta_s$ with $\tau_0=0$.  Processed measurements are represented by $\mf{C}\mf{x}[k]$ with some constant matrix $\mf{C}\in\mathbb{R}^{{n_\mf{z}}\times {n_\mf{x}}}$ with $(\mf{A},\mf{C})$ observable. Each method has a different perception-latency in $\{\Delta^1,\dots,\Delta^D\}$. These latencies are multiples of the sampling period $\Delta_s$ of the sensors leading to a frame-skipping technique. Hence, if method $p_k\in\{1,\dots,D\}$ is chosen at $t=\tau_k$, a new measurement $\mf{z}[k] = \mf{C}\mf{x}[k] + \mf{v}[k]$ is available at $t=\tau_k + \Delta^{p_k}$ where $\mf{v}[k]$ is a noise for the accuracy of the perception method. 

\RevOne{The  perception-latency/accuracy for the perception method is modeled by the covariance matrix $\mf{R}^{p_k}=\cov\{\mf{v}[k]\}$. Typically, increasing the computing time $\Delta^{p_k}$ leads to a decrease in $\mf{R}^{p_k}$ and an improvement in estimation precision. However, we do not make any assumptions about the specific perception-latency and quality model. Instead, for the sake of flexibility, a nominal $\mf{R}^{p_k}$ is expected to be estimated based on the performance of the perception methods for the particular application of interest.}

The main issue we address in this work is to obtain an estimate of the target state $\mf{x}(t)$ at any time $t\geq 0$ and choose which perception method to use at each time slot $[\tau_k,\tau_{k+1})$ provided some performance measure is optimized. In order to provide a concrete definition of the performance measure used in this work, consider the following definitions.

\begin{defn}(Perception schedules)
A perception schedule $p$ is a sequence of perception methods, i.e. $p=\{p_k\}_{k=0}^{\normalfont{\textsf{len}}(p)-1}$ with $p_k\in\{1,\dots,D\}$ for some $\len(p)\in\mathbb{N}\cup\{\infty\}$.
\end{defn}
\begin{defn}(Attention of a perception schedule)
\label{def:attention} Let $\mathcal{I}\subset\mathbb{R}_+$ be an interval with $\sup(\mathcal{I})-\inf(\mathcal{I})>\max\{\Delta^1,\dots,\Delta^D\}$. Thus, the attention $\mfs{att}(p;\mathcal{I})$ of a perception schedule $p$ for $\mathcal{I}$ corresponds to the amount of processed measurements generated by $p$ in such interval.

\end{defn}
Thus, consider the following problem:
\begin{problem}[Perception scheduling]\label{pron:latency_problem}
Design a causal estimator which picks a perception method $p_k$ at each $t=\tau_k$ and produces an estimate $\hat{\mf{x}}(t)$ of $\mf{x}(t)$ with $\hat{\mf{P}}(t):={\cov\{\mf{x}(t)-\hat{\mf{x}}(t)\}}$ using only information prior to the instant $t$. Moreover, the estimator output must minimize 
\begin{equation}
\label{eq:cost}
\begin{aligned}
    \mathcal{J}(\hat{\mf{x}},p) &= \frac{1}{T_f}\left(\int_0^{T_f} \tr\left(\hat{\mf{P}}(t)\right)\text{d}t\right)+ \frac{{\lambda_\alpha}}{T_f}\left(\sum_{k=0}^\alpha r^{p_k}\right)
    \end{aligned}
\end{equation}
for some window of interest $[0,T_f]$, where $\alpha:=\mfs{att}(p;[0,T_f])$, $\lambda_\alpha>0$ and $r^{p_k}\geq 0$ are additional penalties assigned to each perception method.
\end{problem}
\begin{rem}
\label{rem:cost}
\RevAll{The cost function in \eqref{eq:cost} is used to capture the perception latency-accuracy trade-off as explained in the following: the integral term measures the mean expected squared error of $\hat{\mf{x}}(t)$ by evaluating $\mathbb{E}{(\hat{\mf{x}}(t)-\mf{x}(t))^\top(\hat{\mf{x}}(t)-\mf{x}(t))} \equiv \tr(\hat{\mf{P}}(t))$. Thus, the integral term in \eqref{eq:cost} represents the quality of the estimation $\hat{\mf{x}}(t)$. The summation term in \eqref{eq:cost} does not depend on the estimations $\hat{\mf{x}}(t)$ but only on the penalties $r^{p_k}$. If $r^{p_k}=1$ for all $k=0,1,\dots$, this term penalizes the attention of $p$. In some scenarios, it is beneficial to keep this quantity small due to energy consumption. Similarly, non-skipped frames in vision-based perception require an exchange of information between a sensor and the computing unit, and a sensor sampling request. Therefore, processing fewer measurements is desirable for energy efficiency and to minimize the use of I/O buses in the system.}
\end{rem}

\begin{rem}
\label{rem:cpu}
The latency $\Delta^{p_k}$ is usually not exclusively used to process sensor information, but the method also frees the computing unit for an interval of length $(1-f^{p_k})\Delta^{p_k}$ with $f^{p_k}\in(0,1)$. Thus, the CPU load in the interval $[0,T_f]$ can be expressed as $(1/T_f)\sum_{k=0}^\alpha f^{p_k}\Delta^{p_k}$. Therefore, penalties $r^{p_k} = f^{p_k}\Delta^{p_k}$ in \eqref{eq:cost} penalize CPU load.
\end{rem}
\RevTwo{The concepts of attention and CPU load, as discussed in the context of vision-based perception, share similarities with revisit and dwell-times in radar-based perception. Revisit time in a radar corresponds to the interval between consecutive detections, while dwell-time represents the time it takes for the sensor to generate a detection \cite{narykov2013}. Both times have an impact on detection accuracy and energy consumption. Thus, revisit and dwell-times play similar roles to the inverse of attention and CPU load.}

We propose PLATE as a strategy to compute the optimal schedule $p$ and the optimal estimations $\mf{\hat{x}}(t)$ in the sense of Problem \ref{pron:latency_problem}. Causality in Problem \ref{pron:latency_problem} is required in a sequential TT context. Thus, PLATE works as a causal estimator as is summarized in Algorithm \ref{algo:basic_loop} and is comprised by the combination of an estimation stage and a perception-latency scheduling policy.

\AtBeginEnvironment{algorithm}{\let\textnormal\ttfamily}
\begin{algorithm}
\DontPrintSemicolon
\label{algo:basic_loop}
\SetAlgoLined
$\tau_s\assign 0, k\assign 0$\\
\SetKwRepeat{Repeat}{At}{end}
\Repeat(\ sampling event $t=\tau_s$ {\bf do}){}{
\textbf{Estimation stage}: Use $\mf{x}_0$ and $\{\mf{z}[0],\dots,\mf{z}[k-1]\}$ to compute $\hat{\mf{x}}[k]$.\\
Read raw data from sensors. \\
\textbf{Perception latency scheduling}: Decide which perception method $p_k\in\{1,\dots,D\}$ to use. \\
Wait $\Delta^{p_k}$ units of time until the perception-latency has elapsed to produce $\mf{z}[k]$, i.e. $\tau_s\assign \tau_s + \Delta^{p_{k}}$.\\
 $k\assign k+1$\\
 }
 \caption{PLATE loop}
\end{algorithm}

\section{Perception-latency aware estimation}
\label{sec:estimation}

In this section, we establish the structure of the estimation stage of PLATE. To do so, and in order to study the latency-precision trade-off under a cost of the form \eqref{eq:cost}, it is useful to study an equivalent model of \eqref{eq:system_sde} as a sampled-data system. The following result follows from \cite[Section 4.5.2]{Soderstrom}:
\begin{proposition}
\label{th:discrete_system}
Consider a perception schedule $p$. Hence, the solution $\mf{x}(t)$ of \eqref{eq:system_sde} satisfy:
\begin{subequations}
\label{eq:discrete_sde}
    \begin{align}
    \mf{x}(t) &= \mf{A}_d(t-\tau_k)\mf{x}[k] + \mf{w}_d(t) \label{eq:discrete_sde_t}\\
    \mf{x}[k+1]&=\phantom{d.} \mf{A}_d(\Delta^{p_k})\mf{x}[k]+ \mf{w}_d[k] \label{eq:discrete_sde_k}
    \end{align}
\end{subequations}
for $t\in[\tau_k,\tau_{k+1}), k\geq0$ where $\mf{A}_d(t-\tau_k) := \exp(\mf{A}(t-\tau_k))$ and $\mf{w}_d(t)$ normally distributed with $\cov\{\mf{w}_d(t)\}$ given as
\begin{equation}
\label{eq:covariance}
\begin{aligned}
\mf{W}_d(t-\tau_k)&=\int_0^{t-\tau_k}\mf{A}_d(\tau)\mf{BWB}^\top\mf{A}_d(\tau)^\top\text{\normalfont d} \tau
\end{aligned}
\end{equation}
\end{proposition}
\RevFour{The discrete-time nature of the perception process is made explicit by \eqref{eq:discrete_sde}, which is equivalent to \eqref{eq:system_sde}. As a result, we adopt \eqref{eq:discrete_sde} as the motion model throughout the manuscript.}

The following result shows the computation of an optimal estimation $\hat{\mf{x}}(t)$ of ${\mf{x}}(t)$ provided an optimal schedule $p$.
\begin{theorem}
\label{th:kalman}
Let $p$ be the optimal schedule for Problem \ref{pron:latency_problem}. Thus, the optimal estimation for \eqref{eq:system_sde} at $t\in[\tau_k,\tau_{k+1})$ is given by the conditional mean $\mathbb{E}\left\{\mf{x}(t)\ \big|\ \mf{z}[0],\dots,\mf{z}[k-1]\right\}$ and can be computed as: 
\begin{equation}
\label{eq:estimate_time}
\begin{aligned}
\hat{\mf{x}}(t) &=  \mf{A}_d(t-\tau_k)\mf{\hat{x}}[k]\\
\hat{\mf{P}}(t) &= \cov\{\mf{x}(t)-\hat{\mf{x}}(t)\} \\&= \mf{A}_d(t-\tau_k)\hat{\mf{P}}[k]\mf{A}_d(t-\tau_k)^\top + \mf{W}_d(t-\tau_k)
\end{aligned}
\end{equation} for $t\in[\tau_k,\tau_{k+1})$ with $\mf{\hat{x}}[k], \hat{\mf{P}}[k]$ generated according to
\begin{equation}
\label{eq:kalman}
    \begin{aligned}
    \hat{\mf{x}}[0]&=\mf{x}_0, \ \hat{\mf{P}}[0] = \mf{P}_0\\
    \mf{L}[k]&=\mf{A}_d(\Delta^{p_k})\hat{\mf{P}}[k]\mf{C}^\top\left(\mf{C}\hat{\mf{P}}[k]\mf{C}^\top+\mf{R}^{p_k}\right)^{-1} \\
    \hat{\mf{x}}[k+1] &= \mf{A}_d(\Delta^{p_k})\hat{\mf{x}}[k]+\mf{L}[k]\left(\mf{z}[k]-\mf{C}\hat{\mf{x}}[k]\right)\\
    \hat{\mf{P}}[k+1] &=\left(\mf{A}_d(\Delta^{p_k})-\mf{L}[k]\mf{C}\right)\hat{\mf{P}}[k]\left(\mf{A}_d(\Delta^{p_k})-\mf{L}[k]\mf{C}\right)^\top\\&+\mf{L}[k]\mf{R}^{p_k}\mf{L}[k]^\top+\mf{W}_d(\Delta^{p_k})\\
    \end{aligned}
\end{equation}
\end{theorem}
\begin{pf}
The proof can be found in \ref{app:kalman}.
\end{pf}

\section{Perception latency scheduling}
\label{sec:scheduling}

Now that the estimation stage of PLATE has been established, we turn our attention to obtaining the optimal scheduling policy, \RevFour{minimizing \eqref{eq:cost}}. In contrast to the estimations $\hat{\mf{x}}(t)$ which may take arbitrary values in $\mathbb{R}^{n_{\mf{x}}}$, the perception schedule $p$ is of discrete nature. 

\RevFour{One possible way to obtain the optimal schedule that solves Problem \ref{pron:latency_problem} is to use Algorithm \ref{algo:dyn_prog} by calling $\texttt{dynProg}(0,\hat{\mf{P}}[0],T_f)$. This algorithm is essentially an exhaustive search for the optimal schedule, organized as a dynamic programming recursive algorithm.}

\RevFour{Figure \ref{fig:tree} illustrates how our dynamic programming algorithm works. The algorithm starts at time $\tau^+=0$ when calling $\texttt{dynProg}(0,\hat{\mf{P}}[0],T_f)$ and tries out each of the $D$ possible perception methods. For each method, it computes a cost-to-arrive $J$ (line 6 of Algorithm \ref{algo:dyn_prog}) and checks whether the current schedule length $\tau^+$ is less than the desired length $T_f$. If $\tau^+ < T_f$, the algorithm must continue exploring other scheduling options that may extend the current schedule. This process is represented by the black nodes in Figure \ref{fig:tree}. On the other hand, if $\tau^+ \geq T_f$, the algorithm has found a valid schedule that covers the entire time window $[0, T_f]$. This is represented by the white nodes in the figure.}

\RevFour{At each step of the algorithm, a recursive call to the function is made using the current schedule length $\tau^+$ and the updated covariance state $\hat{\mf{P}}[k+1]$. The recursion continues until the algorithm reaches the end of the time window $T_f$. At that point, the algorithm returns the optimal schedule decision $p_k$ and the final cost for that schedule.}

The correctness and complexity of the algorithm are established in the following result:

\begin{figure}
    \centering
    \includegraphics[width=0.45\textwidth]{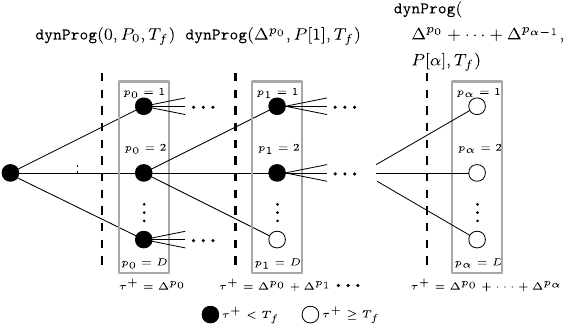}
    \caption{Graphical depiction of Algorithm \ref{algo:dyn_prog}. At each step, all scheduling method options are explored in a recursive fashion with a maximum depth until the condition $\sum_{k=0}^{\alpha}\Delta{p_k}\geq T_f$ where $\alpha=\mfs{att}(p;[0,T_f])$ is reached.}
    \label{fig:tree}
\end{figure}

\begin{proposition}
\label{prop:worst_case}
Calling $\{p,J\}\leftarrow\texttt{dynProg}(0,\hat{\mf{P}}[0],T_f)$ and computing $\hat{\mf{x}}(t)$ using the PLATE structure in \eqref{eq:estimate_time} for such $p$, results in the optimal estimations, schedule and cost $\hat{\mf{x}}, p, \mathcal{J}(\hat{\mf{x}},p)\equiv J$ for Problem \ref{pron:latency_problem} with worst case complexity given by $O\left(D^{\alpha_{\max}}\right)$ where $\alpha_{\max}:=\lfloor T_f/\min\{\Delta^1,\dots,\Delta^D\}\rfloor$.
\end{proposition}
\begin{pf}
The proof can be found in \ref{app:worst_case}.
\end{pf}

\AtBeginEnvironment{algorithm}{\let\textnormal\ttfamily}
\begin{algorithm}[ht]
\DontPrintSemicolon
\label{algo:dyn_prog}
\SetAlgoLined
 \KwData{$\tau,\hat{\mf{P}}[k],T_f$}
\KwResult{$p,J$}

$J\leftarrow \infty$

$p\leftarrow \varnothing$

\For{$\rho \in \{1,\dots,D\}$}
{
$\tau^+\leftarrow \tau + \Delta^{\rho}$

$p^+\leftarrow \varnothing$

$\begin{aligned}
J_\rho&\leftarrow \frac{1}{T_f}\Bigg(\lambda_\alpha r^{\rho} + \int_\tau^{\min(\tau^+,T_f)}\tr(\mf{W}_d(t-\tau))\text{\normalfont d}t+\\ &\left.\int_\tau^{\min(\tau^+,T_f)} \tr(\mf{A}_d(t-\tau)\hat{\mf{P}}[k]\mf{A}_d(t-\tau)^\top)\text{\normalfont d}t\right)\\
\end{aligned}
$

\If{$\tau^+<T_f$}
{
Compute $\hat{\mf{P}}[k+1]$ with \eqref{eq:kalman} for $\rho$ and $\hat{\mf{P}}[k]$

$\{p^+,J^+\}\leftarrow $dynProg$(\tau^+,\hat{\mf{P}}[k+1],T_f)$

$J_\rho\leftarrow J_\rho+J^+$
}

\If{$J_\rho<J$}
{
$J\leftarrow J_\rho$

Append $\rho$ to the start of $p^+$ and assign the result to $p$
}

}

 \caption{dynProg}
\end{algorithm}

The previous result evidences some important complications of Problem \ref{pron:latency_problem}. First, as shown in Proposition \ref{algo:dyn_prog}, the complexity of the exact solution in Algorithm \ref{algo:dyn_prog} increases exponentially as $T_f$ increases. This is not surprising, since the discrete nature of the perception schedule $p$ suggests that the problem is subject to a combinatorial explosion. Moreover, due to the transient behaviour of $\hat{\mf{P}}(t)$, choosing $p_k$ just to minimize a one step ahead of the cost in line 6 of \texttt{dynProg} may not be sufficient to find an exact solution for the problem without having to explore future scheduling method options over the window $[0,T_f]$. Some performance improvements can be made to Algorithm \ref{algo:dyn_prog} such as applying a branch-and-bound technique \cite[Chapter 2.3.3]{bertsekas2000} or general approximations or heuristics \cite[Chapter 6]{bertsekas2000}. However, seeking for tailor-made approximate solution and studying its performance gap with respect to the optimal is more appropriate for a practical implementation. 
\subsection{Quantized covariance approach}
\label{sec:quant}

We provide an approximate alternative to Algorithm \ref{algo:dyn_prog} in order to make it computationally feasible for use in PLATE. This approximation is sub-optimal, but can be made as close to optimal as desired, jeopardizing the computational complexity. The key idea is to observe that as the window $[0,T_f]$ is traversed during Algorithm \ref{algo:dyn_prog}, the set of possible states of the covariance $\hat{\mf{P}}[k]$ increase exponentially in size as well. This is evident from Figure \ref{fig:tree} which shows how the nodes at each level of the exploration tree increase exponentially as all possible combinations of perception methods are tested. However, if the amount of nodes at each level of the tree in Figure \ref{fig:tree} is managed to keep a bound regardless of $T_f$, the complexity of the algorithm can be reduced. This can be performed by grouping similar values of $\hat{\mf{P}}[k]$ into a single element of $\mathbb{R}^{ n_{\mf{x}}\times  n_{\mf{x}}}$ through quantization. The result is a compressed exploration graph with a non-growing number of nodes per level. 

The quantization procedure for the covariance values is described as follows. Consider a compact set $\mathcal{B}_0\subset \mathbb{R}^{n_{\mf{x}}\times n_{\mf{x}}}$ containing only positive semi-definite matrices, characterized by a bound $B_0>0$ such that $\|\hat{\mf{P}}[k]\|_F\leq B_0, \forall \hat{\mf{P}}[k]\in\mathcal{B}_0$. Now, given some $\delta>0$, partition $\mathcal{B}_0$ into $Q_0(\delta)\in\mathbb{N}$ non-overlapping regions $\mathcal{B}_1,\dots,\mathcal{B}_{Q_0(\delta)}$ such that $\sup_{\hat{\mf{P}}'[k],\hat{\mf{P}}''[k]\in\mathcal{B}_{q}}\|\hat{\mf{P}}'[k]-\hat{\mf{P}}''[k]\|_F\leq \delta, \forall q\in\{1,\dots,Q_0(\delta)\}$ and $\bigcup_{q=1}^{Q_0(\delta)}\mathcal{B}_q= \mathcal{B}_0$. Moreover, pick a representative $\hat{\mf{P}}_q\in\mathcal{B}_q$ as an identifier for all other $\hat{\mf{P}}[k]\in\mathcal{B}_q$. Finally, let the quantization function $\mathcal{Q}:\mathcal{B}_0\to\{\hat{\mf{P}}_1,\dots,\hat{\mf{P}}_{Q_0(\delta)}\}$ which takes any $\hat{\mf{P}}[k]\in\mathcal{B}_0$ and maps it to the $\hat{\mf{P}}_q$ such that $\hat{\mf{P}}[k]\in\mathcal{B}_q$. Note that we do not require to use a uniform quantization procedure. In fact, for practical purposes it is convenient to use a non-uniform quantization scheme in which $Q_0(\delta)$ points $\hat{\mf{P}}_1,\dots,\hat{\mf{P}}_{Q_0(\delta)}\in\mathcal{B}_0$ are provided instead, e.g. sampled from $\mathcal{B}_0$, from which the maximum distance $\delta>0$ is obtained.

The idea is to track how the identifiers $\hat{\mf{P}}_q$ with $q\in\{1,\dots,Q_0(\delta)\}$ are related between them using the PLATE estimation stage evolution in \eqref{eq:kalman}.
To do so, a weighted directed graph $(\mathcal{V},\mathcal{E}, \rho)$ is constructed with vertex set  $\mathcal{V}=\{1,\dots,Q_0(\delta)\}$. Moreover, each edge $e=(i,j)\in\mathcal{E}$ corresponds to a connection between $i,j\in\{1,\dots,Q_0(\delta)\}$ when $\hat{\mf{P}}_j=\mathcal{Q}(\hat{\mf{P}}[k+1])$ with $\hat{\mf{P}}[k+1]$ computed using \eqref{eq:kalman} for $\hat{\mf{P}}[k]=\hat{\mf{P}}_i$ and a weight function $\rho:\mathcal{E}\to\{1,\dots,D\}$ with $\rho(e)=p_k$. 

However, it might be the case that $\hat{\mf{P}}[k+1]$ lies outside of $\mathcal{B}_0$ for some $p_k$ and $\hat{\mf{P}}[k]=\hat{\mf{P}}_q$, $q\in\{1,\dots,Q_0(\delta)\}$. To account for these cases, we apply the steps  described in Algorithm \ref{algo:expansion} adding new states outside $\mathcal{B}_0$ as required. The result is a graph $\mathcal{G}$ with ${Q}(\delta)\geq Q_0(\delta)$ states where $\hat{\mf{P}}_q\in\mathcal{B}, \forall q\in\{1,\dots,Q(\delta)\}$ and $\mathcal{B}$ is a region of $\mathbb{R}^{n_{\mf{x}}\times n_{\mf{x}}}$ with $\mathcal{B}_0\subset\mathcal{B}$, and its corresponding bound $\|\hat{\mf{P}}[k]\|_F\leq B, \forall \hat{\mf{P}}[k]\in\mathcal{B}$. When Algorithm \ref{algo:expansion} finishes, it is ensured that if $\hat{\mf{P}}[0]\in\mathcal{B}_0$, then the quantized covariance trajectories will be contained in $\mathcal{B}$ for any perception schedule.

\AtBeginEnvironment{algorithm}{\let\textnormal\ttfamily}
\begin{algorithm}[!ht]
\DontPrintSemicolon
\label{algo:expansion}
\SetAlgoLined
 \KwData{$\delta, \mathcal{B}_0$ with $\|\hat{\mf{P}}[0]\|\leq B_0$}
\KwResult{$\mathcal{B}, \mathcal{G}$}

Quantize $\mathcal{B}_0$ into $Q_0(\delta)$ patches $\{\mathcal{B}_1,\dots,\mathcal{B}_{Q_0(\delta)}\}$ with identifiers $\{\hat{\mf{P}}_1,\dots,\hat{\mf{P}}_{Q_0(\delta)}\}$

$q\leftarrow 0$

$\mathcal{V} \leftarrow \{1,\dots,Q_0(\delta)\}$

$\mathcal{E} \leftarrow \varnothing$

$\mathcal{B}\leftarrow \mathcal{B}_0$

${Q}(\delta)\leftarrow {Q}_0(\delta)$

\While{$q\leq Q(\delta)$}
{
    \For{$p \in\{1,\dots,D\}$}
    {
        Compute $\hat{\mf{P}}[k+1]$ from \eqref{eq:kalman} using $\hat{\mf{P}}[k]=\hat{\mf{P}}_q$ for $p_k = p$
        
        \If{$\hat{\mf{P}}_{q'}\neq\mathcal{Q}(\hat{\mf{P}}[k+1])$ for any $q'\in\{1,\dots,Q(\delta)\}$}
        {
            $Q(\delta)\leftarrow Q(\delta)+1$
            
            $\mathcal{V}\leftarrow \mathcal{V} \cup \{Q(\delta)\}$
            
            Create a new patch $\mathcal{B}_{Q(\delta)}$ and identifier $\hat{\mf{P}}_{Q(\delta)}$ such that $\hat{\mf{P}}_{Q(\delta)},\mathcal{Q}(\hat{\mf{P}}[k+1])\in\mathcal{B}_{Q(\delta)}$ and $\sup_{\hat{\mf{P}},\hat{\mf{P}}'\in\mathcal{B}_{Q(\delta)}}\|\hat{\mf{P}}-\hat{\mf{P}}'\|=\delta$
            
            $\mathcal{B}\leftarrow \mathcal{B}\cup \mathcal{B}_{Q(\delta)}$
        }
        Add edge $e=(q,q')$ to $\mathcal{E}$ with weight $\rho(e) = p$
    }
    
    $q\leftarrow q + 1$
}
Construct $\mathcal{G}$ using $\mathcal{V}$, $\mathcal{E}$ and its weights from line 16.
\caption{expandB}
\end{algorithm}

It is beneficial for practical purposes, to estimate how $Q(\delta)$ grows with respect to  original size of $\mathcal{B}_0$ as a result of Algorithm \ref{algo:expansion}. 
The following result ensures that Algorithm \ref{algo:expansion} finishes with finite $Q(\delta)$. In addition, we provide an explicit upper bound for $B$, from which a worst case of $Q(\delta)$ can be computed depending on the actual quantization scheme.

\begin{proposition}
\label{eq:finite_time}
Algorithm \ref{algo:expansion} finishes for any  compact set $\mathcal{B}_0\subset \mathbb{R}^{n_\mf{x}\times n_\mf{x}}$ and $\delta>0$. In addition, the resulting bound $B$ for $\mathcal{B}$ complies
\begin{equation}
    \label{eq:switched_bound}
\begin{aligned}
    B\leq B_s:=\sqrt{n_{\mf{x}}}\left(\frac{\lambda_{\max}(\bm{\Omega})}{\lambda_{\min}(\bm{\Omega})}\right)\left(B_0+\frac{\overline{G}}{1-\gamma}\right)
    \end{aligned}
\end{equation}
where $\bm{\Omega}\in\mathbb{R}^{n_{\mf{x}}\times n_{\mf{x}}}$ and $ \mf{Y}^i\in\mathbb{R}^{n_{\mf{x}}\times n_{\mf{z}}}, i\in\{1,\dots,D\}$ satisfy the following Linear Matrix Inequality (LMI):
\begin{equation}
\label{eq:lmi}
\begin{aligned}
&\begin{bmatrix}
\gamma \bm{\Omega} & (\bm{\Omega} \mf{A}_d(\Delta^i)-\mf{Y}^i\mf{C})^\top \\ (\bm{\Omega} \mf{A}_d(\Delta^i)-\mf{Y}^i\mf{C}) & \bm{\Omega}
\end{bmatrix} \succeq 0,\\& \quad \quad i=1,\dots,D
\end{aligned}
\end{equation}
for some $0<\gamma<1$. Moreover, $\lambda_{\min}(\bm{\Omega}),\lambda_{\max}(\bm{\Omega})$ are the minimum and maximum eigenvalues of $\bm{\Omega}$ respectively and  $\overline{G}:=\max_{i\in\{1,\dots,D\}}\|(\mf{L}^{p_k})\mf{R}^{p_k}(\mf{L}^{p_k})^\top+\mf{W}_d(\Delta^{p_k})\|_F $ with $\mf{L}^{p_k}=\bm{\Omega}^{-1}\mf{Y}^{p_k}$.
\end{proposition}
\begin{pf}
The proof can be found in \ref{ap:finite_time}
\end{pf}

\begin{figure}
    \centering
    \includegraphics[width=0.45\textwidth]{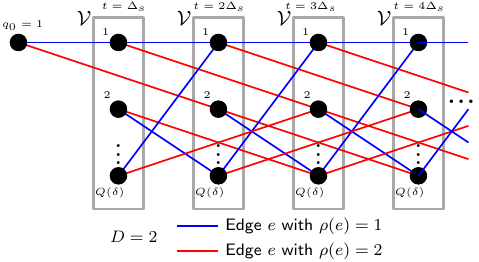}
    \caption{Transition graph example for covariance evolution using $\mathcal{G}$. In this example $D=2$. Each node labeled in $1,\dots,Q(\delta)$ is connected to other nodes through one of the two perception decisions, either one or two steps ahead since $\Delta^1=\Delta_s, \Delta^2=2\Delta_s$ for this example.}
    \label{fig:trellis}
\end{figure}

Using the graph $\mathcal{G}$, the evolution of covariance states in $\mathcal{V}$ can be tracked from an initial condition $q_0\in\mathcal{V}$ as time advances in $[0,T_f]$. This can be visualized using a transition diagram, as shown in Figure \ref{fig:trellis} which depicts an example with $D=2$ and $\Delta^1=\Delta_{s}, \Delta^2=2\Delta_{s}$. Here, at $t=0$ the initial state is $q_0$ corresponding to $\hat{\mf{P}}_{q_0}=\mathcal{Q}(P[0])$. We consider a discrete time $t=\ell\Delta_{s}$ as integer multiples of the minimum sampling interval. As $\ell$ increases, edges $e\in\mathcal{E}$ connect states between time steps, with different perception decisions $\rho(e)$. Note that, under this perspective, number of states at each time step is maintained constant. Hence, a dynamic programming algorithm for finite-state deterministic systems can be applied in order to obtain the optimal schedule $p$ \cite[Page 64]{bertsekas2000}. In this context, the optimal $p$ corresponds to the shortest route starting at $q_0$, which traverses the transition diagram of $\mathcal{G}$ until it reaches any state at stage $\ell=\lfloor T_f/\Delta_{s}\rfloor$ with distance measured by $\mathcal{J}(\mf{\hat{x}},p)$. 

The concrete steps required to run this dynamic programming solution, called \texttt{qDP}, are described in Algorithm \ref{algo:approx_2} in \ref{ap:algo}. Moreover, the sense in which we verify correctness of the dynamic programming as from Algorithm \ref{algo:approx_2} and its complexity is established in the following:

\begin{proposition}
\label{prop:optimal_approx}
Consider the \texttt{qDP} algorithm and assume that the evolution of $\hat{\mf{P}}[k]$ is constrained to evolve according to the structure in $\mathcal{G}$ and that $\hat{\mf{P}}[0]\in\{\hat{\mf{P}}_1,\dots,\hat{\mf{P}}_{Q(\delta)}\}$. Thus, \texttt{qDP} obtains the optimal value of \eqref{eq:cost} for such constrained trajectories. Moreover, the worst-case of complexity of \texttt{qDP} is $O(\alpha_{\max}Q(\delta)D)$ where $\alpha_{\max}=\lfloor T_f/\Delta_{s}\rfloor$.
\end{proposition}
\begin{pf}
The proof can be found in \ref{ap:optimal_approx}.
\end{pf}

As evidenced by the previous result, the complexity of the quantized covariance approach is reduced to something linear in $T_f$. Nonetheless, since this is an approximate solution, there will be a trade-off between complexity and the performance gap of the quantized solution with respect to the true optimal. Note that by decreasing $\delta$, the number of patches $\mathcal{B}_q$ needed to cover $\mathcal{B}$ will increase. However, it is expected that as $\delta\to 0$, the resulting sub-optimal solution perfomance improves. These ideas are formalized in the following result.
\begin{theorem}
\label{th:closedness}
Let $\mathcal{J}$ be the optimal cost for Problem \ref{pron:latency_problem} and $P[0]\in\mathcal{B}_0$. Then, for any $\varepsilon>0$ there exists sufficiently small $\delta>0$ such that
$$
|\mathcal{J} - \mathcal{J}_{\mathcal{Q}}(\delta)|\leq \varepsilon
$$
where $\mathcal{J}_{Q}(\delta)$ is the cost obtained from \texttt{qDP} in \ref{ap:algo} for such $\delta$ and initial condition $\mathcal{Q}(\hat{\mf{P}}[0])\in\mathcal{B}_0$.
\end{theorem}
\begin{pf}
The proof can be found in \ref{app:cost}.
\end{pf}

Henceforth, we have established that the PLATE strategy with estimation stage in \eqref{eq:kalman} and  perception-scheduling policy obtained using \texttt{qDP} is a sub-optimal for Problem  \ref{pron:latency_problem},  with mild computational complexity in terms of the window size $T_f$ and can approximate the optimal solution with arbitrary precision if more computational power is available.

\begin{rem}
\label{rem:quant}
\RevFive{It is important to note that only the perception schedule computation is impacted by quantization, while the actual estimator in \eqref{eq:estimate_time} and \eqref{eq:kalman} is not quantized. Therefore, the quantization step $\delta$ only affects the resulting quality of the perception schedules in terms of the cost \eqref{eq:cost}. We provide an analysis of the asymptotic effect of quantization on the cost in Theorem \ref{th:closedness}. While decreasing $\delta$ is expected to increase the quality of the perception schedules, it is not obvious whether the performance can be made arbitrarily close to the optimal one. This accuracy feature is ensured by Theorem \ref{th:closedness}.} 
\end{rem}

\begin{rem}
\label{rem:particle}
\RevFive{Our approach can be extended to filter structures different than \eqref{eq:kalman} \RevFive{such as a particle filter}, as long as a similar transition graph for the covariance matrix is obtained for its use in Algorithm \ref{algo:expansion}. However, obtaining similar theoretical guarantees as in this work for other filter structures is not a trivial task and require more in-depth analysis.}
\end{rem}

\section{Moving horizon PLATE}
\label{sec:MHS}
One property of the $\texttt{qDP}$ method used in PLATE is that given an initial condition $\hat{\mf{P}}[0]$, a schedule $p$ for the whole time window $[0,T_f]$ is obtained. Thus, it only suffices to compute the schedule at the beginning and traverse it element by element in line 5 of Algorithm \ref{algo:basic_loop} after each perception method is used. However, in a more practical TT scenario, there will be missing measurements as a result of occlusions, or distinguishability problems between targets in a multi-target setting. An appropriate re-detection of the targets may be performed by a maintenance mechanism as widely discussed in the literature. This may alter the quality of the pre-computed perception schedule. 

In addition, the processed measurement quality, represented by the covariances $\{\mf{R}^{1},\dots,\mf{R}^{D}\}$ may not remain fixed in a practical setting. Instead, it is usual to use the  current estimation for the state of the target to improve the quality of subsequent measurements as in \cite{song2019,dong2016,cyber_rnn}.
Moreover, the detection method may include some measure of uncertainty of their current output, e.g. through Bayesian techniques \cite{loquercio2020}, \RevFive{or data-driven methods \cite{fokin2009}}. Thus, an online covariance estimate $\mf{R}[k]$ for the processed measurements may be available and may not be any of the nominal covariances. This discussion motivates to change the perception schedule according to the current state of the system by using \texttt{qDP} as a predictive policy.

To do so, we propose to use moving-horizon scheme for PLATE in the following way. First, we construct the transition graph $\mathcal{G}$ as described in the previous section for nominal covariances $\{\mf{R}^{1},\dots,\mf{R}^{D}\}$. Now, the dynamic programming algorithm \texttt{qDP} is used at $\tau_k$ with initial condition given by the current $q_0=\mathcal{Q}(\hat{\mf{P}}[k])$ to obtain a perception schedule $p'=\texttt{qDP}(q_0,T_f,\mathcal{G})$ for the next window $[\tau_k,\tau_k+T_f]$. As a result, we apply the first perception method of $p'$ during $[\tau_k, \tau_k+\Delta^{p_k})$ and repeat the procedure for the next interval.

Furthermore, every time a new perception output is obtained, the PLATE correction in \eqref{eq:kalman} is computed with the actual processed measurement uncertainty $\mf{R}[k]$ if available, otherwise with its nominal covariance. When there are no measurements, we simply predict the state of the target and its covariance using \eqref{eq:estimate_time}. This procedure is summarized in Algorithm \ref{algo:MHA} which describes the steps that must be performed in line 5 of Algorithm \ref{algo:basic_loop} in order to implement this strategy.

\AtBeginEnvironment{algorithm}{\let\textnormal\ttfamily}
\begin{algorithm}
\DontPrintSemicolon
\label{algo:MHA}
\SetAlgoLined

\tcp{\texttt{\textbf{Offline:}}}

\texttt{Pre-compute $\mathcal{G}$ using Algorithm \ref{algo:expansion}, alternatively pre-compute $\texttt{qDP}(\mathcal{Q}(\hat{\mf{P}}_q, \mathcal{G}, T_f)$ for all $\hat{\mf{P}}_q, q\in\{1,\dots,Q(\delta)\}$.}

\tcp{\texttt{\textbf{Online:}}}

\eIf{There is new processed measurement}
{
    Update $\hat{\mf{P}}[k]$ using \eqref{eq:kalman} for the previous scheduling decision $p_{k-1}$ and its resulting  perception uncertainty $\mf{R}[k-1]$.\\

}
{
Predict $\hat{\mf{P}}[k]$ using \eqref{eq:estimate_time} for the previous scheduling decision $p_{k-1}$.
}
$p', \_ \leftarrow \texttt{qDP}(\mathcal{Q}(\hat{\mf{P}}[k]), \mathcal{G}, T_f)$

$p_k\leftarrow$ first element of $p'$

 \caption{Moving-horizon PLATE}
\end{algorithm}

Moreover, the latency of executing \texttt{qDP} at each $\tau_k$ can be considered negligible if two of the following approaches is used. First, since $\hat{\mf{P}}[k]$ is quantized through $\mathcal{Q}(\bullet)$ there are only $Q(\delta)$ possible outcomes for which their resulting perception method obtained through \texttt{qDP} can be pre-computed offline. On the other hand, similarly to what is suggested in \cite{aldana2020}, taking $\hat{\mf{P}}[k]$ and converting it to a perception method in $\{1,\dots,D\}$ is a classification problem with a low dimensional input. Thus, the perception decision can be learned by a classifier.

\begin{rem}
\changes{
Comparing our approach with previous work, note that the perception scheduling policy as in Algorithm \ref{algo:MHA} acts as a frame-skipping technique similar to \cite{guan2018,casares2011,luo2021}. However, unlike prior frame-skipping approaches which are based on heuristic rules, we provide an actual performance guarantees for our sub-optimal perception schedules as given in Theorem \ref{th:closedness}. On the other hand, note that unlike related multi-sensor scheduling approaches in the literature as in \cite{ss_fast,ss_fast2,ss_greedy,ss_periodic_filter,ss_periodic_ih,ss_pruning,isa2},  Algorithm \ref{algo:approx_2} can be extended for different cost function expressions by modifying line 12 of Algorithm \ref{algo:approx}, which makes our proposal more versatile. }
\end{rem}

\begin{rem}
The moving horizon PLATE proposal allows us to change the system parameters online as in line 3 of Algorithm \ref{algo:MHA}, where we use an uncertainty $\mf{R}[k-1]$ which need not to be any of the nominal ones for the perception methods. This poses an important advantage with respect to the philosophy of some multi-sensor scheduling works such as \cite{ss_periodic_filter,ss_periodic_ih} which rely on periodic schedules or other strategies that are highly dependant on the problem structure.
\end{rem}

\section{Numerical experiments}
\label{sec:simu}
In order to evaluate PLATE we consider the following scenario. First, let a target with sate $\mf{x} = [x,v_x,y,v_y]^\top$ describing its position and velocity on the plane. Let camera images available every $\Delta_{s}=1/30$ seconds, coinciding with the usual frame rate of a camera. In addition, the position $[x,y]^\top$ is measured through a detection process from which there are $D=2$ available perception configuration. First, a lightweight detector with latency $\Delta^1=3\Delta_{s}$ and nominal covariance $\mf{R}^1=\diag(0.5,0.5)$. Second, a more accurate detector with latency $\Delta^2=9\Delta_{s}$ and nominal covariance $\mf{R}^2=\diag(0.05,0.05)$. Motivated by Remark \ref{rem:cpu} we consider CPU loads of $f^1=0.5$ and $f^2=0.8$ resulting in penalties $r^1 = 0.5\Delta^{1}$ and $r^2=0.8\Delta^2$. The system matrices in this scenario are:
$$
\mf{A} = \begin{bmatrix}
0&1&0&0 \\
0&0&0&0 \\
0&0&0&1 \\
0&0&0&0 \\
\end{bmatrix}, 
\mf{C} = \begin{bmatrix}
1&0&0&0\\
0&0&1&0
\end{bmatrix}
$$
In addition, let the process covariance $\mf{W}=\diag(0.5,0.5)$. This double integrator target model is often used for generic targets in the literature. In the following, we evaluate the different aspects of our proposals for this setting.

\subsection{Numerical covariance bound estimation}
\begin{figure}
    \centering
    \includegraphics[width=0.45\textwidth]{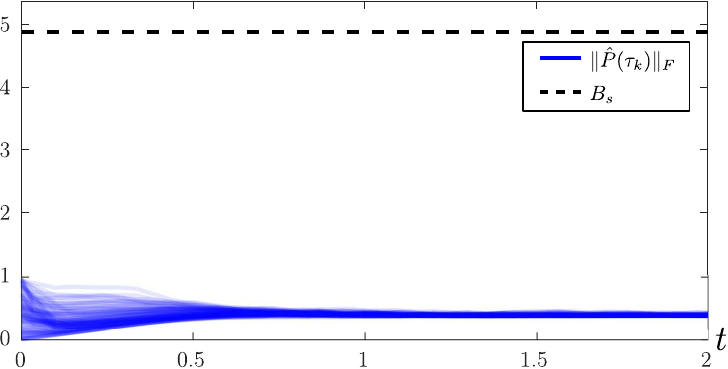}
    \caption{Covariance norm evolution, showing the bound $B_s$ as well as $\|\hat{\mf{P}}(\tau_k)\|$ for PLATE with 100 random initial conditions $\hat{\mf{P}}[0]$ and schedules $p$ with $\|\hat{\mf{P}}[0]\|_F\leq B_0=1$.}
    \label{fig:kalman}
\end{figure}

We start by building a transition graph $\mathcal{G}$. To do so, we set $\mathcal{B}_0\in\mathbb{R}^{n_\mf{x}\times n_\mf{x}}$ as the set of all positive definite matrices $\mf{P}$ with $\|\mf{P}\|_F\leq \mf{B}_0:=1$ and use the mechanism in Algorithm \ref{algo:expansion}. In order to depict the result in Proposition \ref{eq:finite_time}, we explicitly obtain a bound for $\mathcal{B}$. First, note that the LMI \eqref{eq:lmi} has a solution for $\gamma=0.98$ as
$$
\mf{Y}^1= \begin{bmatrix} 0.031&0\\0.037&0\\0&0.031\\0&0.037\end{bmatrix}, \mf{Y}^2= \begin{bmatrix} 0.122&0\\0.137&0\\0&0.122\\0&0.137\end{bmatrix}
$$
with a resulting bound \eqref{eq:switched_bound} of $B_s=4.922$. Figure \ref{fig:kalman} depicts how given initial conditions $P[0]\in\mathcal{B}_0$ and a randomly generated perception schedule, the magnitude of the PLATE covariance complies $\|\hat{\mf{P}}[k]\|_F\leq B_s, \forall k\geq 0$. 

\subsection{Cost comparison using \texttt{qDP}}
\label{sec:histogram}
Now, in order to depict Theorem \ref{th:closedness} through this example, we use different levels of quantization for evaluation. Note that the results presented until now don't require uniform quantization. In fact, in the following examples for simplicity we use non-uniform quantization by sampling a fixed number of points $\hat{\mf{P}}_1,\dots,\hat{\mf{P}}_{Q(\delta)}\in\mathcal{B}_0$ and quantizing any other $\hat{\mf{P}}\in\mathcal{B}_0$ through a nearest neighbor rule under the $\|\bullet\|_F$ norm.  Next, we expand $\mathcal{G}$ through Algorithm \ref{algo:expansion} to obtain $\mathcal{B}$. \RevFive{We use $Q(\delta)=50,500,5000$ which after this process results  in $\delta=10.45, 4.23, 2.14$ respectively for these arbitrarily selected points, drawn randomly over $\mathcal{B}_0$ with uniform distribution, in order to analyze the impact of the quantization step in the tracking quality.} 

\begin{figure}
    \centering
    \includegraphics[width=0.45\textwidth]{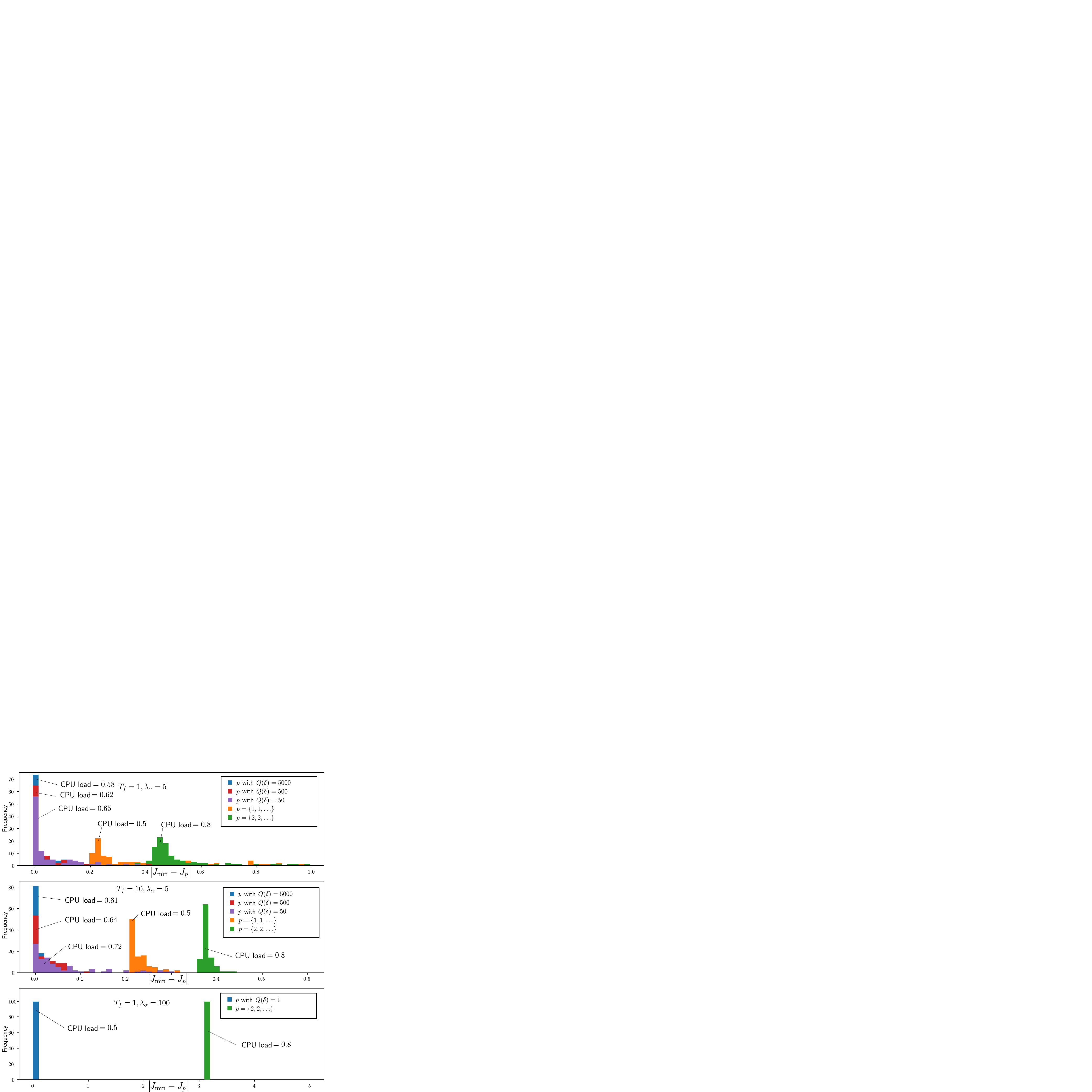}
    \caption{Resulting histograms for the cost difference $|J_{\min}-J_p|$ with the schedule obtained from the approximate dynamic programming approach for $100$ random initial conditions for $P[0]$ as described in Section \ref{sec:histogram}. The parameters $T, \lambda_\alpha$ of the cost function in \eqref{eq:cost} were changed for three different scenarios as well as the number of quantization points $Q(\delta)$. In addition, only the average CPU load for all experiments is shown for convenience.}
    \label{fig:histogram}
\end{figure}

We evaluate \texttt{qDP} for $100$ random samples for $P[0]$ over $\mathcal{B}$ and compute their resulting cost \eqref{eq:cost}. First, consider $T_f=1, \lambda_\alpha=5$. In this case, it is possible to obtain the minimum cost $J_{\min}$ by evaluating all schedules covering the window $[0,T_f]$. We show the results in the first row of Figure \ref{fig:histogram} where the histograms depict the frequency of the distance $|J_{\min}-J_p|$ where $J_p$ is the cost obtained for a schedule $p$. The schedules tested where obtained through \texttt{qDP} for graphs $\mathcal{G}$ using the previously described values of $Q(\delta)$, as well as the static schedules $p=\{1,1,\dots\}$ and $p=\{2,2,\dots\}$. The results show that as the quanitization gets finer, the cost concentrate more and more towards the minimum, i.e. $|J_{\min}-J_p|$ becomes smaller. In addition, it is observed that even with a coarse discretization $Q(\delta)=50$, the cost of \texttt{qDP} is almost always less than the cost for the static schedules. Moreover, the average CPU load seems to decrease as well as the discretization gets finer. A similar behaviour is obtained when using $T_f=10, \lambda_\alpha=100$, as depicted in the second row of Figure \ref{fig:histogram}. In this case, $J_{\min}$ is computed for $10000$ different randomly selected perception schedules aiming to approximate the otherwise intractable exhaustive search for the true optimal cost. 

Another interesting experiment is obtained by increasing $\lambda_\alpha=15$ such that the cost of the covariance is negligible to the one of the penalties $r^{p_k}$ in \eqref{eq:cost}. In this case, $J_{\min}$ is directly the static schedule $p=\{1,1,\dots\}$ which is the schedule with less CPU load. In addition, since the particular values of covariance $\hat{\mf{P}}[k]$ have little impact on the final cost, it only suffices to use a single point $Q(\delta)=1$ in the discretization of the covariance space. Hence, the result of \texttt{qDP} is the static schedule $p=\{1,1,\dots\}$ regardless of the initial condition $\hat{\mf{P}}[0]$ as shown in the third row of Figure \ref{fig:histogram}.

\subsection{Moving-horizon PLATE scheduling}
\label{sec:movPLATE}
\begin{figure}
    \centering
\includegraphics[width=0.45\textwidth]{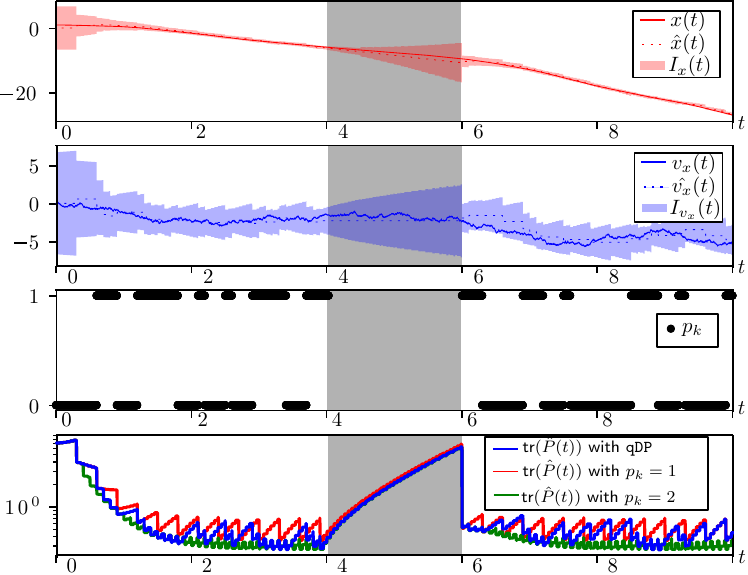}
    \caption{Behaviour of the target model \eqref{eq:system_sde} and the output of PLATE using the moving-horizon scheduling with $T_f=10$ and $Q(\delta)=5000$ when an occlusion occurs for $t\in[4,6]$. Moreover, $I_x(t)$ and $I_{v_x}(t)$ represent confidence intervals around $\hat{x}(t)$ and $\hat{v}_x(t)$ of 3 times their standard deviation.}
    \label{fig:occlusion}
\end{figure}

\RevFive{We simulate \eqref{eq:system_sde} for $\mf{x}_0=0, P[0]= 4I$ using the Euler-Maruyama discretization with time step $\Delta t= 10^{-3}$ and run the loop in Algorithm \ref{algo:basic_loop} under the moving-horizon scheduling of Algorithm \ref{algo:MHA} where \texttt{qDP} is configured similarly as in the previous examples, with $T_f=10, \lambda_\alpha=5, Q(\delta)=5000, B_0=5$ (Algorithm \ref{algo:expansion}).} Figure \ref{fig:occlusion} depicts the time evolution of $\mf{x}(t)$ as well as the PLATE estimate $\hat{\mf{x}}(t)$, where we show only $x$ and $v_x$ for convenience, since $y$ and $v_y$ behave similarly. In addition, we show confidence intervals of 3 times the standard deviation for each coordinate obtained from $\hat{\mf{P}}(t)$. In addition, we show the scheduling decision $p_k$ at each time. In order to compare the cost in this experiment we show $\tr(\hat{\mf{P}}(t))$ for the moving-horizon scheduling, as well as for the static schedules. It is worth noting that the best quality measured with $\tr(\hat{\mf{P}}(t))$ is obtained by the static schedule with $p_k=2$. However, it has the highest CPU load of 0.8. The moving-horizon PLATE strategy manages to have an intermediate quality between the two static schedules, with a CPU load of 0.65 for this experiment. Thus, PLATE manages to obtain a better trade-off between quality and resource usage. In addition, an occlusion is simulated in the interval $t\in[4,6]$ which shows how uncertainty increases during this period, but recovers once new measurements arrive. 

\section{Evaluation on real data}
\label{sec:realdata}
\RevAll{In this section, we evaluate PLATE on a real target tracking task. We use the MOT16 benchmark from \cite{MOT16} since it is part of a public standard data-set, widely used to evaluate target tracking algorithms. In the simplest setting, the task is to take images from a monocular camera and track a target of interest across multiple frames. As usual, and for the sake of simplicity, we are interested in tracking the pixel position of the geometric center of a bounding box surrounding the target. For this example, we model the pixel position with a two-dimensional single integrator with fixed $n_\mf{x}=n_\mf{z}=n_\mf{w}=2$, $\mf{A} = \mf{0}\in\mathbb{R}^{2\times 2}, \mf{B} = \mf{C}= \mf{I}\in\mathbb{R}^{2 \times 2}$.}

\SecondRound{In the following, we describe a target tracking pipeline that consists of a perception stage composed of three different target detection methods based on neural networks, each exhibiting varying latency and accuracy performance, and a scheduling stage responsible for deciding which of these methods to use at each time. The objective of this section is to demonstrate that PLATE is the best choice for the scheduling stage. For this purpose, we compare the tracking performance of PLATE with two other approaches: 1) a standard neural network-based method, which involves using the perception methods separately with no scheduling, 2) event-triggered approaches for frame-skipping.}

\subsection{Evaluation framework}
\label{sec:realdataPipeline}

\RevAll{The MOT16 data-set includes 7 video sequences, each with labeled bounding-boxes for different moving targets. We manually selected several targets for each sequence. We recorded their ground-truth positions and the corresponding portion of the sequence, which we refer to as target tracks from now on. We chose targets that provided longer experiments, in order to obtain more comprehensive results. Our evaluation data consists of 50 target tracks across all 7 video sequences, whose ground-truth IDs are shown in Table \ref{tab:dataset} as a reference. Note that for all video sequences the frame rate is set to 30 frames per second except for MOT16-05 and MOT16-13 with frame rates of 14 and 25 frames per second respectively. As a result, target tracks for MOT16-13 and MOT16-05 (15 tracks in total) were used exclusively to estimate the nominal detection covariance for each perception method as described later, and the covariance $\mf{W}$ in \eqref{eq:system_sde}. This process is called training in Table \ref{tab:dataset}. The rest of the 35 target tracks contained in the remaining 5 video sequences were used to evaluate the online performance of PLATE.} \SecondRound{
From this point onward, all experiments (both for our proposal and other proposals used for comparison) were conducted on a PC equipped with an Intel Core i7-8700, along with the aid of an NVIDIA GeForce GTX 1080 Ti graphics card.}
\begin{table}
\centering
\RevAll{
\scalebox{0.9}{
\begin{tabular}{||c l c c||} 
 \hline
 Name & Target IDs & fps& Usage\\ [0.5ex] 
 \hline
   MOT16-05 & 2,37,41,124,128&14& {\sf T}\\
  MOT16-13 &13,14,29,30,76,77,79,81,95,124& 25&{\sf T}\\ 
  \hline
  MOT16-02 & 3,19,20,32,33,34,35 & 30& {\sf E}\\
  MOT16-04 & 2,67,70,80,83,98,99,100 &30& {\sf E}\\
  MOT16-09 & 12,13,19,20,23&30& {\sf E}\\
  MOT16-10 & 1,2,4,7,18,24 &30& {\sf E}\\
  MOT16-11 & 1,3,4,5,6,11,25,27,29&30& {\sf E}\\ 
 \hline
\end{tabular}}}
\caption{\RevAll{Ground-truth target IDs, as provided by the MOT16 data-set, picked for each video sequence in our evaluation. Frames per second is abbreviated as fps. The last column indicates {\sf T} or {\sf E} if the target tracks in the corresponding video sequence are used for training or for evaluation respectively.}}
\label{tab:dataset}
\end{table}

\RevAll{To evaluate the performance of PLATE, it is necessary to use a setup that includes a bank of perception methods. However, the purpose of our work is not to evaluate the individual performance of these methods, but rather to measure the advantages of PLATE in terms of target state estimation quality and resource usage, such as CPU load. Therefore, we selected well-established perception methods from the public model zoo of Detectron2 \cite{wu2019detectron2}. This repository provides many neural network-based target detection algorithms with varying levels of latency and precision, making it suitable for our work. Specifically, we chose the  \texttt{faster\_rcnn\_R\_50\_FPN\_3x} and \texttt{faster\_rcnn\_R\_101\_C4\_3x} models (hereinafter referred as the fast and slow networks respectively), which represent the clearest examples in the zoo of low quality detection with short latency and good quality detection with long latency, respectively. These models have a reported latency of  $0.038$s and $0.104$s on a standard computing platform, which were verified in our own setup. To further simplify the experiments, for the fast network we sub-sample the input image in a factor of two, reducing the latency to roughly $1/(30\mfs{fps})\approx 0.033s$ which is the inverse of the frame rate. With these two networks we propose 3 perception methods as follows:
\begin{itemize}
\item $p_k=1$:  This method uses the fast network with a latency of $1/(30\mfs{fps})$. This means that this method does not skip any frame and has a CPU load of $100\%$.
\item $p_k=2$: It uses the slow network with a processing latency of $0.104s$. This means that the network is computed over four frames, causing it to skip three frames, processing only the first one. Consequently, the CPU load over these four frames is $\frac{0.104s}{4\mfs{frames}/(30\mfs{fps})} = 78\%$.
\item $p_k=3$: In order to have an option which favours a low CPU load, we use the fast network, but deliberately skip the next 4 frames. This means that the perception quality of this method is the same as $p_k=1$ but with a CPU load of $\frac{1/30\ s}{5\mfs{frames}/(30\mfs{fps})}=20\%$ and longer overall latency.
\end{itemize}}

\begin{figure}
    \centering
\includegraphics[width=0.4\textwidth]{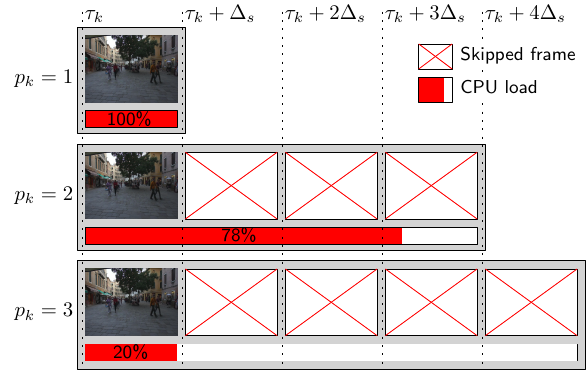}

    \caption{\RevAll{Different features of the three perception methods described in Section \ref{sec:realdataPipeline}. The number of skipped frames as well as the CPU load in each method is depicted.} }
    \label{fig:modes}
\end{figure}

\RevAll{These perception method options are depicted graphically in Figure \ref{fig:modes}.} \RevFive{The previous framework has a disadvantage in terms of memory consumption, as the adaptive strategy requires storing both neural networks, leading to higher memory usage compared to using the perception methods individually. However, it is worth noting that depending on the application, it may be possible to use a single ANN, as demonstrated in \cite{hu2019}, which can offer different latency and quality levels with a fixed amount of memory space.}

\begin{figure}[ht]
    \centering\includegraphics[width=0.4\textwidth]{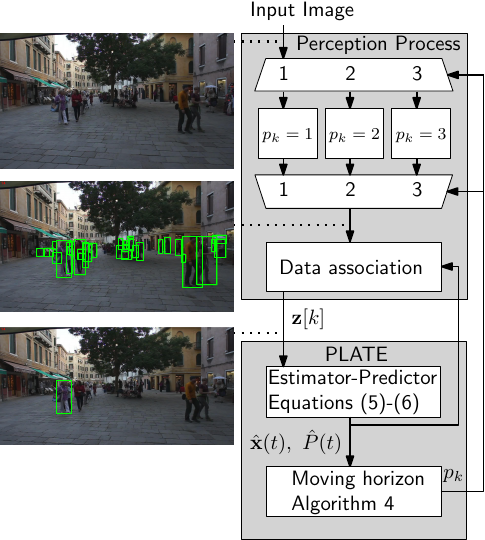}
    
    \caption{\RevAll{Implemented Pipeline for the PLATE evaluation framework described in Section \ref{sec:realdataPipeline}}}
    \label{fig:pipeline}
\end{figure}

\RevTwo{As depicted in Figure \ref{fig:pipeline}, the output of each of the neural networks is a list of bounding boxes for all the detected objects in the image. Therefore, a data association stage is necessary to select the appropriate bounding box for the target object in each experiment. There exist many data association techniques in the literature, such as those based on similarity measures using neural networks \cite{zhou2019} or end-to-end detection and association siamese networks \cite{siamese2016}. These methods are highly dependent on the training procedure, which might have an undesired effect when evaluating the scheduling feature of PLATE by itself. Hence, since the goal of this work is to evaluate PLATE for a fixed perception framework, we use a more standard approach. We compare the predicted position of the target, $\lim_{t\to\tau_k^-}\hat{\mf{x}}(t)$, from \eqref{eq:estimate_time} at the time $\tau_k$ when the image was captured, and use the Hungarian algorithm to select the most suitable bounding box candidate \cite{luo2021}. The computational latency of this procedure is negligible and can be ignored for simplicity in subsequent discussions.  Figure \ref{fig:pipeline} shows that the output of the data association stage is the best bounding box candidate for the target of interest from which a processed position measurement $\mf{z}[k]$ is obtained as its geometric center.}

\RevAll{ In this setting, $\Delta_s=1/30, \Delta^1=\Delta_s, \Delta^2=3\Delta_s, \Delta^3=5\Delta_s$. In addition, the CPU loads are  $f^1=1, f^2=0.78, f^3=0.2$. We estimate the nominal error covariance for each perception method $\mf{R}^1, \mf{R}^2, \mf{R}^3$ by comparing the outputs of each perception method with the ground-truth data across all images in the 15 training target tracks. The resulting covariance matrices are}
\RevAll{$$
\mf{R}^1 =\diag(13.12^2, 25.87^2), \mf{R}^2=\diag(9.94^2, 17.06^2), \mf{R}^3=\mf{R}^1 
$$}
\RevAll{The matrix $\mf{W}$ was estimated using the training target tracks as well, following a standard parameter estimation procedure described in \cite{fokin2009}. As described in Section \ref{sec:MHS}, we also aim to evaluate PLATE when an online estimation of the current covariance $\mf{R}[k]$ is available besides the nominal covariances. As a result, we adapt the standard parameter estimation in \cite{fokin2009} for this setting, using the resulting performance of previous detections compared to the estimation $\hat{\mf{x}}(t)$. Let $\mathcal{I}^{p_k}[k]=\{\ell\leq k : p_\ell=p_k, k-\ell\leq N_w\}$ be the set of all discrete instants $\ell$ corresponding to the moments $\tau_\ell$ in which the perception method $p_k$ was used prior to the current $\tau_k$, in a moving window picked here of $N_w=10$ samples. Thus, if $p_k$ is to be picked at $t=\tau_k$ one can estimate the current covariance for such method as:
\begin{equation}
\label{eq:adaptiveR}
\mf{R}[k] = \frac{1}{N_w}\sum_{\ell\in\mathcal{I}^{p_k}[k]}\mf{e}[\ell]\mf{e}[\ell]^\top - \mf{C}\hat{\mf{P}}^-(\tau_k)\mf{C}^\top
\end{equation}
where $\mf{e}[\ell] = \mf{C}\hat{\mf{x}}[\ell]-\mf{z}[\ell]$ and $\hat{\mf{P}}^-(\tau_k) = \lim_{t\to\tau^-}\hat{\mf{P}}(t)$ from \eqref{eq:estimate_time}.}

\RevAll{The PLATE module, depicted in Figure \ref{fig:pipeline}, executes the estimator-predictor equations specified in \eqref{eq:estimate_time} and \eqref{eq:kalman}, and implements the scheduler algorithm presented in Algorithm \ref{algo:MHA}. This module employs a moving-horizon PLATE with $T_f=10s$ and utilizes pre-computed schedules for each quantized covariance value $\hat{\mf{P}}_q$, which are based on nominal $\mf{R}^1, \mf{R}^2,$ and $\mf{R}^3$. When a new processed measurement $\mf{z}[k]$ is obtained, the pre-computed values for line 7 in Algorithm \ref{algo:MHA} for all possible $\hat{\mf{P}}[k]$ in a compact set enable the computation of the new scheduling decision $p_k$ with negligible computing latency. \RevFour{Additionally, the computation time required for \eqref{eq:estimate_time} and \eqref{eq:kalman} in this example's state space dimensions is negligible compared to the latencies $\Delta^1, \Delta^2,$ and $\Delta^3$.} The values of $Q(\delta)$ will be varied between $50,500,5000$ as in previous examples with $\mathcal{B}_0$ as the set of all positive definite matrices $\mf{P}$ with $\|\mf{P}\|_F\leq \mf{B}_0=20$}.

\begin{rem}
\label{rem:penalty}
\RevAll{The cost function \eqref{eq:cost} was set using $\lambda_\alpha=1$ and 
\begin{equation}
\label{eq:penalty}
r^{p_k} = \lambda_{\mfs{load}}f^{p_k}\Delta^{p_k}+\lambda_{\mfs{att}}.
\end{equation}
The term $\lambda_{\mfs{load}}f^{p_k}\Delta^{p_k}$ penalizes the CPU load, while $\lambda_{\mfs{att}}$ penalizes the attention, weighted by $\lambda_{\mfs{load}}, \lambda_{\mfs{att}}$.} \RevTwo{Minimizing these objectives while providing the best possible accuracy are conflicting goals, but can be balanced by adjusting $\lambda_{\mfs{load}},\lambda_{\mfs{att}}$. If $\lambda_{\mfs{load}}=\lambda_{\mfs{att}}=0$, the offline construction of $p_k$ results in $p_k=1$ for every $\hat{\mf{P}}_q$, maximizing the attention regardless of the quantization step. Conversely, we verified numerically that for $\lambda_{\mfs{load}}=1$ and $\lambda_{\mfs{att}}>15$, $p_k=3$ for all $\hat{\mf{P}}_q$, minimizing the attention, regardless of the quantization step as well. For other values of $\lambda_{\mfs{load}}, \lambda_{\mfs{att}}$, the schedule $p_k$ changes according to the current $\hat{\mf{P}}_q$, achieving a balance between these objectives.} 
\end{rem}

\subsection{Performance of the implemented pipeline}
\label{sec:discussion}

\RevAll{In this section, we evaluate the proposed pipeline on the 35 evaluation target tracks. The performance evaluation metric used is the Mean Squared Error (MSE) of $\hat{\mf{x}}(k\Delta_s), k=0,1,\dots$ compared to the ground-truth data available in the MOT16 data-set. Additionally, the CPU load for all experiments is recorded, as well as the attention measured as the percentage of non-skipped frames compared to the total number of frames. 
} \SecondRound{To illustrate the trade-off between the previously described metrics, a combined cost is computed as $(\text{MSE})+\lambda_\mfs{load}(\text{CPU load}) + \lambda_\mfs{att}(\text{Attention})$ as an sampled version of the cost \eqref{eq:cost} under penalties \eqref{eq:penalty}.}

\RevAll{In the following, we used $\lambda_{\mfs{load}}=\lambda_{\mfs{att}}=1/2$ in \eqref{eq:penalty} which results in a trade-off between accuracy, CPU load and attention, producing different $p_k$ through time for PLATE. The results are shown in Table \ref{tab:results} where we compare the performance of the proposal under different values of ${Q}(\delta)$ as well as using nominal $\mf{R}[k]\in\{\mf{R}^1,\mf{R}^2,\mf{R}^3\}$ or adaptive $\mf{R}[k]$ in \eqref{eq:adaptiveR}. } 

\RevAll{There are three main conclusions that can be obtained from these results. First, note that using adaptive $\mf{R}[k]$ produces an improvement with respect to the nominal case, particularly with low $Q(\delta)$ where an improvement of $(35.76-32.78)/35.76\approx 8\%$ in MSE is obtained when $Q(\delta)=50$. Second, consistent with Theorem \ref{th:closedness}, increasing $Q(\delta)$ improves the MSE as well, with diminishing returns as $Q(\delta)$ increases. \SecondRound{Third, the CPU load and attention remain roughly the same meaning that high values of $Q(\delta)$ might not be needed in practice to maintain reasonable values for these objectives.}}

\SecondRound{Now, we compare PLATE with other ideas in the literature. As a baseline for the comparison we used a framework which does not use scheduling. This is, the neural network method to be used for perception is chosen at the beginning, and remains fixed at all times. The results are shown in the first three rows of Table \ref{tab:results1}. Note that the configuration in row 1) of Table \ref{tab:results1} (fixed $p_k=1$) obtains the best MSE among all experiments at the expense of high CPU load and attention. In addition, row 3) of Table \ref{tab:results1} (fixed $p_k=3$) produces the best results in terms of CPU load and attention at the expense of high MSE. However, recall from Remark \ref{rem:penalty} that with appropriate configuration of the penalties \eqref{eq:penalty}, PLATE can produce either fixed $p_k=1$ or $p_k=3$ and obtain the best performance for MSE, CPU load or attention by separate.}

\SecondRound{Moreover, we evaluated frame-skipping techniques \cite{guan2018,casares2011,luo2021} which rely on an event-triggered condition to determine whether a new frame needs processing. To evaluate a similar approach, we decided at each new frame whether to use $p_k=1$ or skip the frame entirely. Send-on-delta strategies are commonly employed, where a frame is processed if $\tr(\hat{\mf{P}}(k\Delta_s)) \geq \delta_{\mfs{ET}}$ with a threshold $\delta_{\mfs{ET}}>0$. The results for different $\delta_{\mfs{ET}}$ are shown in rows 4) and 5) of Table \ref{tab:results1}.}

\SecondRound{Finally, row 6) of Table \ref{tab:results1} shows the best configuration for PLATE as obtained in Table \ref{tab:results}. For comparison, note that while fixed $p_k=2$ yields a small MSE, CPU load and attention values remain high. In contrast, PLATE in row 6) of Table \ref{tab:results1} also has a small MSE with the advantage of reducing the CPU load from $78\%$ to $43.5\%$ and attention from $75\%$ to $43.4\%$,  when compared to fixing $p_k=2$.}

\SecondRound{The event-triggered approach in row 4) of Table \ref{tab:results1} shows a small improvement in the MSE can be obtained when compared to PLATE in row 6) of Table \ref{tab:results1}. Despite this, the CPU load and attention values are considerably bigger for the event triggered alternative. More concretely, PLATE in row 6) of Table \ref{tab:results1} improves the CPU load from $71.2\%$ to $43.5\%$ (a relative improvement of $38.9\%$) when compared to the event triggered approach in row 4) of Table \ref{tab:results1}, even with a similar MSE performance. This trade-off is clear since PLATE obtains the best performance among all options as ilustrated with the combined cost in the last column of Table \ref{tab:results1}.}

\SecondRound{Hence, using PLATE with appropriate configuration in \eqref{eq:penalty} can obtain a good trade-off between accuracy, CPU load and attention, with improved performance with respect to static perception configurations. In addition, PLATE provides a clear connection via the cost function \eqref{eq:cost} and the penalty \eqref{eq:penalty} to achieve in each situation different behaviors,
 prioritizing CPU load and attention or MSE. }

\begin{rem}\label{rem:parameters}
\SecondRound{The performance of the proposal is highly dependent on the parameters $Q(\delta)$, $\lambda_{\mfs{load}}$, and $\lambda_{\mfs{att}}$, as evident from the results presented in Tables \ref{tab:results} and \ref{tab:results1}. It is important to note that the optimal parameter selection generally varies based on the specific application and target dynamics under consideration. However, as discussed in this section, conducting several experimental tests can help determine the most suitable parameters for the user. Our experiments demonstrate a range of potential performance values obtained by varying the parameter $Q(\delta)$, allowing for selection based on specific application requirements and resource availability. In addition, regarding the penalties $\lambda_{\mfs{load}}$, and $\lambda_{\mfs{att}}$, these values should reflect the application specifications. Still, it is beneficial to identify parameter sets for these penalties that exhibit similar behavior to the ones described in Remark \ref{rem:penalty}. This knowledge becomes particularly valuable when the user needs to make a trade-off between accuracy and resource usage. More principled design rules are not trivial to obtain and will be considered in our future work.}
\end{rem}

\begin{table*}
\centering
\RevAll{
\scalebox{0.8}{
\begin{tabular}{||c c c c c c c ||} 
 \hline
  & $\mf{R}[k]$ &$Q(\delta)$ & MSE [\mfs{px}] & CPU load [\mfs{\%}] & Attention [\mfs{\%}] & Combined cost \\ 
  \hline
1) & Nominal &$50$ &35.76 &44.5& 44.1& 80.06\\
2) & Adaptive &$50$ &32.78 &44.0 & 43.9& 76.73\\
3) & Nominal &$500$ &32.14 &44.1 & 43.9& 76.14\\
4) & Adaptive &$500$   &32.02 & 43.8& 43.6& 75.72\\
5) & Nominal &$5000$ &31.54 &43.9 & 43.6& 75.29\\
6) & Adaptive &$5000$  &\textbf{30.04} & \textbf{43.5}& \textbf{43.4}& \textbf{73.49}\\
 \hline
\end{tabular}}}
\caption{\RevAll{Evaluation of the implemented pipeline for PLATE on MOT16 data as described in Section \ref{sec:realdata}. MSE stands for Mean-Squared-Error with respect to ground-thruth data. The attention column corresponds to the percentage of non-skipped frames with respect to the total number of frames. The penalty in \eqref{eq:penalty} is configured with $\lambda_{\mfs{load}}=\lambda_{\mfs{att}}=1/2$}. \SecondRound{Moreover, to illustrate the trade-off between the previously described metrics, a combined cost is computed as $(\text{MSE})+\lambda_\mfs{load}(\text{CPU load}) + \lambda_\mfs{att}(\text{Attention})$ as an sampled version of the cost \eqref{eq:cost} under penalties \eqref{eq:penalty}.}}
\label{tab:results}
\end{table*}

\begin{table*}
\centering
\RevAll{
\scalebox{0.8}{
\begin{tabular}{||c l l c c c c||} 
 \hline
  &Method &Configuration & MSE [\mfs{px}] & CPU load [\mfs{\%}] & Attention [\mfs{\%}] & Combined cost \\ 
  \hline
  1) & {No scheduling} & fixed $p_k=1$ &{\bf 19.61} &100 & 100 & 119.61\\
  2) & {No scheduling} &fixed $p_k=2$ &29.05 &78 & 75 & 105.55\\
  3) & {No scheduling} &fixed $p_k=3$ &73.73 &{\bf 20} & {\bf 20} & 93.73\\
  4) & {Event-triggered} &$\delta_{\mfs{ET}}=5$ &25.04 &71.2 &69.2 & 95.24\\
  5) & {Event-triggered} &$\delta_{\mfs{ET}}=50$ &64.74 & 35.9& 35.2 & 100.29\\
6) & {\bf PLATE}  &$Q(\delta)=5000$, Adaptive $\mf{R}[k]$  &{\bf 30.04} & {\bf 43.5}& {\bf 43.4}& {\bf 73.49}\\
 \hline
\end{tabular}}}
\caption{\SecondRound{Evaluation of the implemented pipeline. The evaluation metrics are the MSE, CPU load and attention as well as the combined cost taking into account all previous metrics weighted by $\lambda_{\mfs{load}}=\lambda_{\mfs{att}}=1/2$. The methods under evaluation are: 1) No scheduling with fixed $p_k=1$, corresponding to using only the fast neural network \texttt{faster\_rcnn\_R\_50\_FPN\_3} at all times without additional frame skipping. 2) No scheduling with fixed $p_k=2$ corresponding to using only the slow neural network \texttt{faster\_rcnn\_R\_101\_C4\_3x} at all times without additional frame skipping. 3) No scheduling with fixed $p_k=3$ corresponding to using only the fast neural network \texttt{faster\_rcnn\_R\_50\_FPN\_3} followed by 4 skipped frames repeatedly. 4) and 5) Event triggered approach with different $\delta_{\mfs{ET}}$. 6) The best configuration for PLATE as obtained from Table \ref{tab:results}.}}
\label{tab:results1}
\end{table*}

\section{Conclusions}
\RevAll{In this work, we introduced PLATE as a perception-latency aware estimator for target tracking applications. PLATE leverages a bank of perception methods with different latency-precision trade-offs to adaptively select the best method for the current estimation task. The proposed algorithm allows for the skipping of input frames, which reduces CPU load and resource usage while still maintaining high tracking accuracy. Unlike other frame skipping techniques, PLATE's algorithm is based on a formal dynamic programming argument rather than heuristics. We found that while the exact solution of the problem is subject to a combinatorial explosion, an approximate solution can be obtained with efficient computational complexity. We evaluated PLATE using both simulations and real-world data sets, and found that it outperforms other state-of-the-art approaches in terms of both tracking accuracy and computational efficiency.} \RevAll{Our current approach is limited to linear target motion models, and extending the estimator to nonlinear models using particle filtering or data-driven techniques is challenging and left for future work.} 
\section{Acknowledgements}
This work was supported via projects PID2021-124137OB-I00 and TED2021-130224B-I00
funded by MCIN/AEI/10.13039/501100011033, by ERDF A way of making Europe and by the
European Union NextGenerationEU/PRTR, by the Gobierno de Aragón under Project DGA T45-23R, by the Universidad de Zaragoza and Banco Santander, by the Consejo Nacional de Ciencia y Tecnología (CONACYT-Mexico) with grant number 739841.

\appendix
\section{Quantized covariance method}
\label{ap:algo}

The concrete steps to perform the dynamic programming algorithm described in Section \ref{sec:quant} are shown in Algorithm \ref{algo:approx_2}. First, auxiliary matrices $\mf{M}_Q,\mf{M}_P,\mf{M}_J$ are computed from Algorithm \ref{algo:approx} given an initial state $q_0$. In line 6 of Algorithm \ref{algo:approx} time steps up to $\alpha_{\max}$ are traversed. Moreover, in line 7 all possible states $q$ at time $\ell$ are evaluated as well. Furthermore, for each of these states $q$, in line 8 all perception options $\rho$ are checked. In this way, Algorithm \ref{algo:approx} tracks optimal routes where $[\mf{M}_J]_{q,\ell}$ stores the best cost-to-arrive from $q_0$ to $q$ after $\ell$ time steps. Moreover, $[\mf{M}_Q]_{q',\ell}$ stores the best state $q$ connected to $q'$ at time step $\ell$, where the perception decision is stored in $[\mf{M}_P]_{q',\ell}$ defining how many stages separate both states as well. These matrices are used in Algorithm \ref{algo:approx_2} to trace-back the optimal scheduling.

\AtBeginEnvironment{algorithm}{\let\textnormal\ttfamily}
\begin{algorithm}[ht]
\DontPrintSemicolon
\label{algo:approx_2}
\SetAlgoLined
 \KwData{$q_0,T_f,\mathcal{G}$ computed from Algorithm \ref{algo:expansion}}
\KwResult{$p, J$}

$\{\mf{M}_Q,\mf{M}_P,\mf{M}_J\}\leftarrow$\texttt{qDPMatrices} $(q_0,T_f,\mathcal{G})$

Set $J$ and $q$ to the best cost and state at the last column of $\mf{M}_J$

$p\leftarrow \varnothing$

\tcp{Use $\mf{M}_P$ to trace back the optimal schedule ending at state $q$}

$\ell\leftarrow \lfloor T_f/\Delta_{s}\rfloor$

\While{$\ell>0$}
{
    $\rho\leftarrow[\mf{M}_P]_{q,\ell}$

    Append $\rho$ at the start of $p$ 
    
    $q\leftarrow [\mf{M}_Q]_{q, \ell}$
    
    $\ell\leftarrow \ell-\Delta^\rho/\Delta_s$

}
 \caption{qDP}
\end{algorithm}

\AtBeginEnvironment{algorithm}{\let\textnormal\ttfamily}
\begin{algorithm}[ht!]
\DontPrintSemicolon
\label{algo:approx}
\SetAlgoLined
 \KwData{$q_0,T_f, \mathcal{G}$}
\KwResult{$\mf{M}_Q,\mf{M}_P,\mf{M}_J$}

$\alpha_{\max}\leftarrow \lfloor T_f/\Delta_{s}\rfloor$

$\mf{M}_Q \leftarrow[0]\in\{0,1,\dots,Q(\delta)\}^{Q(\delta)\times (\alpha_{\max}+1)}$
\tcp{$[\mf{M}_Q]_{q',\ell}$: best state $q$ connected to $q'$ at step $\ell$}

$\mf{M}_P \leftarrow[1]\in\{0,1,\dots,D\}^{D\times (\alpha_{\max}+1)}$
\tcp{$[\mf{M}_P]_{q',\ell}$: Perception connecting best $q$ connected to $q'$ at step $\ell$}

$\mf{M}_J \leftarrow [\infty]\in\bar{\mathbb{R}}_{\text{\fontsize{1}{0}$\geq \!\!0$}}^{Q(\delta)\times (\alpha_{\max}+1)}$ \tcp{$[\mf{M}_J]_{q,\ell}$: best cost from $q_0$ to $q$ in $\ell$ steps}

$[\mf{M}_J]_{q_0,0}\leftarrow 0$

\For{$\ell\in\{1,\dots,\alpha_{\max}\}$}
{
    \For{$q\in \{1,\dots,Q(\delta)\}$}
    {
        
        \For{$\rho\in\{1,\dots,D\}$}
        {
            $q'\leftarrow$ state connected to $q$ through edge with weight $\rho$
            
            $\tau\leftarrow (\ell-1)\Delta_{s}$
            
            $\tau^+\leftarrow \tau + \Delta^{\rho}$
            
            \small
            $\begin{aligned}
            J&\leftarrow \frac{1}{T_f}\Bigg(\lambda_\alpha r^{\rho} + \int_\tau^{\min(\tau^+,T_f)}\mf{W}_d(t-\tau)\text{\normalfont d}t+ \\ &\left.\int_\tau^{\min(\tau^+,T_f)} \tr(\mf{A}_d(t-\tau)\hat{\mf{P}}_q\mf{A}_d(t-\tau)^\top)\text{\normalfont d}t\right)\\
            \end{aligned}$\normalsize
            
            \tcp{$[\mf{M}_J]_{q,\ell}+J$ is the cost-to-arrive from $q_0$ to $q'$ in $\ell+\rho$ steps}
            
            \If{$[\mf{M}_J]_{q,\ell}+J<[\mf{M}_J]_{q',\ell+\rho}$}
            {
                $[\mf{M}_J]_{q',\ell+\rho}\leftarrow [\mf{M}_J]_{q,\ell}$
                
                $[\mf{M}_Q]_{q',\ell+\rho}\leftarrow q$
                
                $[\mf{M}_P]_{q',\ell+\rho}\leftarrow \rho$
            }
        }
    }
}

 \caption{qDPMatrices}
\end{algorithm}

\section{Auxiliary results }
\label{app:aux}
\begin{proposition}\label{prop:lp_norm} Let $\bm{\lambda}=[\lambda_1,\dots,\lambda_n]^\top\in\mathbb{R}^n$ and define $\|\bm{\lambda}\|_p = \left(\sum_{i=1}^n|\lambda_i|^{p}\right)^{1/p}$. Then, with $0<r<s$, the following inequalities are satisfied:
\begin{enumerate}[a)]
    \item\label{prop:lp_norm_rs} \cite[Theorem 16, Page 26]{hardy} $\|\bm{\lambda}\|_r\leq n^{\frac{1}{r}-\frac{1}{s}}\|\bm{\lambda}\|_s$.
    \item\label{prop:lp_norm_sr} \cite[Theorem 19, Page 28]{hardy} $\|\bm{\lambda}\|_s\leq \|\bm{\lambda}\|_r$.
\end{enumerate}
\end{proposition}
\begin{lemma}
\label{le:bounds}
Let $\mf{M}_1,\mf{M}_2\in\mathbb{R}^{n\times n}$ be positive definite matrices such that $\mf{M}_1\preceq \mf{M}_2$. Then, the following inequalities are satisfied:
\begin{enumerate}[a)]
    \item $\tr(\mf{M}_1)\leq \tr(\mf{M}_2)$.
    \item $\|\mf{M}_1\|_F\leq \sqrt{n}\|\mf{M}_2\|_F$
\end{enumerate}
\end{lemma}
\begin{pf}
First, note that $\mf{M}_2-\mf{M}_1$ is positive semi-definite. Then, use the spectral Theorem \cite[Theorem 2.5.6]{horn} to conclude that the  eigenvalues $\{\lambda_i\}_{i=1}^n$ of $\mf{M}_2-\mf{M}_1$ are all different non-negative real numbers. Thus, $\tr(\mf{M}_2)-\tr(\mf{M}_1)=\tr(\mf{M}_2-\mf{M}_1)=\sum_{i=1}^n\lambda_i\geq 0$, where the relation between the trace and the sum of the eigenvalues was used in the last step \cite[Page 50]{horn} completing the proof for a). Similarly, $\mf{M}_1,\mf{M}_2$ have non-negative eigenvalues too. Let $\bm{\lambda}_1,\bm{\lambda}_2\in\mathbb{R}^n$ be vectors containing the eigenvalues of $\mf{M}_1,\mf{M}_2$ respectively. Thus, item a) reads $\|\bm{\lambda}_1\|_1\leq \|\bm{\lambda}_2\|_1$ using the notation of Proposition \ref{prop:lp_norm} in \ref{app:aux}. Proposition \ref{prop:lp_norm}-a) implies $\|\bm{\lambda}_2\|_1\leq \sqrt{n}\|\bm{\lambda}_2\|_2$. Proposition \ref{prop:lp_norm}-b) implies $\|\bm{\lambda}_1\|_2\leq\|\bm{\lambda}_1\|_1$. Thus, $\|\bm{\lambda}_1\|_2\leq \sqrt{n}\|\bm{\lambda}_2\|_2$. Now, recall that $\|\bm{\lambda}_1\|_2 = \sqrt{\sum_{i=1}^n \lambda_i(\mf{M}_1)^2}\equiv \|\mf{M}_1\|_F$ where $\lambda_i(\mf{M}_1)$ are the eigenvalues of $\mf{M}_1$ and similarly for $\mf{M}_2$ \cite[Page 342]{horn}. Thus, item b) follows.
\end{pf}

\begin{lemma}
\label{lem:kron_def}
Let $\mf{M}_1,\mf{M}_2\in\mathbb{R}^{n\times n}$ be symmetric matrices satisfying $\mf{M}_1\preceq \mf{M}_2$. Then, $(\mf{M}_1\otimes \mf{M}_1)\preceq (\mf{M}_2\otimes \mf{M}_2)$.
\begin{pf}
First, recall from \cite[Theorem 7.7.3-(a)]{horn} that $\mf{M}_1\preceq \mf{M}_2$ if and only if $\rho(\mf{M}_2^{-1}\mf{M}_1)\leq 1$ where $\rho(\bullet)$ denotes the spectral radius \cite[Definition 1.2.9]{horn}. Then, $1\geq \rho(\mf{M}_2^{-1}\mf{M}_1)^2= \rho((\mf{M}_2^{-1}\mf{M}_1)\otimes(\mf{M}_2^{-1}\mf{M}_1)) = \rho((\mf{M}_2\otimes \mf{M}_2)^{-1}(\mf{M}_1\otimes \mf{M}_1))$, which implies $(\mf{M}_1\otimes \mf{M}_1)\preceq (\mf{M}_2\otimes \mf{M}_2)$ using \cite[Theorem 7.7.3-(a)]{horn}.
\end{pf}
\end{lemma}

\section{Proofs}

\subsection{Proof of Theorem \ref{th:kalman}}
\label{app:kalman}
First, we show an following auxiliary result.
\begin{lemma}
\label{th:astorm_kalman}
Let the state equation \eqref{eq:discrete_sde} and the measurement model $\mf{z}[k]=\mf{C}\mf{x}[k] + \mf{v}[k]$ available at $t=\tau_{k}+\Delta^{p_k}$ with $\cov\{\mf{v}[k]\} = \mf{R}^{p_k}$ and a fixed latency schedule $p$. Thus, the estimate at time $t\in[\tau_k,\tau_{k+1})$ of $\mf{x}(t)$ based on measurements $\{\mf{z}[0],\dots,\mf{z}[k]\}$ which minimize $\tr(\hat{\mf{P}}(t))$, is given by $\mathbb{E}\{\mf{x}(t)|\mf{z}[0],\dots,\mf{z}[k-1]\}$ and satisfies \eqref{eq:estimate_time}.
\end{lemma}
\begin{pf}
First, consider estimates for $\mf{x}[k]$, given that $\mf{x}[k]$ evolves according to the discrete-time system \eqref{eq:discrete_sde_k}. Moreover, note that the measurement $\mf{z}[k]$ is available at $t=\tau_{k+1}$. Then, \cite[Page 228 - Theorem 4.1]{astrom} implies that the structure in \eqref{eq:kalman}, inherited from a Kalman filter with predictor, satisfies $\hat{\mf{x}}[k+1]\equiv\mathbb{E}\{\mf{x}[k+1]|\mf{z}[0],\dots,\mf{z}[k]\}$. For any other $t\in(\tau_k,\tau_{k+1})$ the measurement $\mf{z}[k]$ is not available. Thus, $\mathbb{E}\{\mf{x}(t)|\mf{z}[0],\dots,\mf{z}[k-1]\}$ can be computed using $\mathbb{E}\{\mf{x}[k]|\mf{z}[0],\dots,\mf{z}[k-1]\}$ through the same structure in \eqref{eq:kalman} but applied to \eqref{eq:discrete_sde_t} with $\mf{C}=0$ resulting in \eqref{eq:estimate_time}. Now that $\hat{\mf{x}}(t)=\mathbb{E}\{\mf{x}(t)|\mf{z}[0],\dots,\mf{z}[k-1]\}$ has been established for any $t\in[\tau_k,\tau_{k+1})$, \cite[Page 228 - Theorem 4.1]{astrom} implies that  $\mf{a}^\top\hat{\mf{P}}(t)\mf{a}$ is minimized for this estimate with arbitrary vector $\mf{a}\in\mathbb{R}^{n_{\mf{x}}}$. This means that for any other estimation $\mf{\hat{x}}'(t)$ of $\mf{x}(t)$ with covariance $\hat{\mf{P}}'(t):=\cov\{\mf{x}(t)-\mf{\hat{x}}'(t)\}$ we have $\mf{a}^\top\hat{\mf{P}}(t)\mf{a}\leq\mf{a}^\top\hat{\mf{P}}'(t)\mf{a}, \forall \mf{a}\in\mathbb{R}^{n_\mf{x}}$. Hence, $\hat{\mf{P}}(t)\preceq\hat{\mf{P}}'(t)$ from which $\tr(\hat{\mf{P}}(t))\leq \tr(\hat{\mf{P}}'(t))$ follows using Lemma \ref{le:bounds} from \ref{app:aux} for any other estimation $\mf{x}'(t)$.
\end{pf}
We are now ready to show Theorem \ref{th:kalman}. First, note that for fixed perception schedule $p$, the term $\sum_{k=0}^\alpha r^{p_k}$ in \eqref{eq:cost} is constant. Moreover, Lemma \ref{th:astorm_kalman} implies that $\mf{\hat{x}}(t)=\mathbb{E}\{\mf{x}(t)|\mf{z}[0],\dots,\mf{z}[k-1]\}$ computed recursively using \eqref{eq:estimate_time} and \eqref{eq:kalman} leads to an optimal trajectory of the signal $\tr(\hat{\mf{P}}(t))$. This means that for any other estimation $\hat{\mf{x}}'(t)$ of $\mf{x}(t)$ with $\hat{\mf{P}}'(t):=\cov\{\mf{x}(t)-\hat{\mf{x}}'(t)\}$ one has $\tr(\hat{\mf{P}}(t))\leq \tr(\hat{\mf{P}}'(t))$. Integrating both sides of the previous inequality leads to conclude that $\int_0^{T_f} \tr(\hat{\mf{P}}(t))\nd t \leq \int_0^{T_f} \tr(\hat{\mf{P}}'(t))\nd t$ which implies $\mathcal{J}(\hat{\mf{x}},p)\leq \mathcal{J}(\hat{\mf{x}}',p)$. Hence, the estimation $\mf{\hat{x}}(t), \forall t\in[0,T_f]$ from \eqref{eq:estimate_time} solves Problem \ref{pron:latency_problem} provided that $p$ is the optimal schedule as well.

\subsection{Proof of Proposition \ref{prop:worst_case}}
\label{app:worst_case}
To show correctness, first note that a direct application of the dynamic programming principle \cite[Page 23]{bertsekas2000} leads to conclude that $p$ as obtained from $\texttt{dynProg}(0,P[0],T_f)$ is optimal for the cost \eqref{eq:cost} whenever the estimations $\{\hat{\mf{x}}(t):t\in[0,T_f]\}$ are of the form \eqref{eq:estimate_time} as required in lines 4 and 6 of \texttt{dynProg}. In addition, Theorem \ref{th:kalman} implies that the optimal estimations are computed from \eqref{eq:estimate_time}, from which optimality of both $\hat{\mf{x}}(t)$ and $p$ using \texttt{dynProg} follows, solving Problem \ref{pron:latency_problem}. For the complexity, note that the largest amount of recursive calls in Algorithm \ref{algo:dyn_prog} is obtained when $p_k=i$ at line 3 of \texttt{dynProg} with $i=\text{argmin}\{\Delta^j\}_{j=1}^D$. In this case, $\lfloor T_f/\min\{\Delta^1,\dots,\Delta^D\} \rfloor$  recursive calls to \texttt{dynprog} are required to cover the whole window $[0,T_f]$, i.e. for $\tau^+\geq T_f$ in line 7 of \texttt{dynProg}. Hence, due to line 3 in \texttt{dynProg}, the total number of recursive calls is at most $D^{\lfloor T_f/\min\{\Delta^1,\dots,\Delta^D\}\rfloor}$.

\subsection{Proof of Proposition \ref{eq:finite_time}}
\label{ap:finite_time}

First we show that Algorithm \ref{algo:expansion} finishes. Note that for any initial condition in a compact set $\mathcal{B}_0$, the result in \cite[Lemma 6.1]{anderson} ensures that the covariance $\hat{\mf{P}}[k]$ a the filter of the form \eqref{eq:kalman} is uniformly bounded from above. This means that there exists a compact set ${\mathcal{B}}\subset \mathbb{R}^{n_\mf{x}\times n_\mf{x}}$ such that $\hat{\mf{P}}[k]\in{\mathcal{B}}, \forall k\geq 0$ and $\hat{\mf{P}}_0\in\mathcal{B}_0$. Existence of such compact set in the quantized setting is ensured as well. The reason is that quantization only adds a disturbance term $\|\bm{\Pi}_\mathcal{Q}[k]\|\leq \delta$ to \eqref{eq:kalman} which can be absorbed into the covariance $W(\Delta^{p_k})$. Now, note that since quantization patches obtained in Algorithm \ref{algo:expansion} are non overlapping and comply $\sup_{\hat{\mf{P}}[k],\hat{\mf{P}}[k]'\in\mathcal{B}_{Q(\delta)}}\|\hat{\mf{P}}[k]-\hat{\mf{P}}[k]'\|=\delta$ thus, there exists a maximum finite number $\bar{Q}(\delta)$ of regions of this kind covering $\mathcal{B}$ due to compactness. Hence, Algorithm \ref{algo:expansion} finishes in at most $\bar{{Q}}(\delta)$ steps.

For the rest of the proposition, the proof outline is described in the following. Note that obtaining an explicit bound for $\hat{\mf{P}}[k]$ in \eqref{eq:kalman} is not trivial due to the nonlinearity of the update equation for the covariance. In addition, the bound in \cite[Lemma 6.1]{anderson} is very overestimated. Hence, we follow a similar idea as in \cite{anderson} which is to study an artifact filter whose bound can be obtained explicitly and use it as a proof tool, appealing to the optimality of PLATE to conclude that a covariance bound of the artifact filter is a bound for PLATE as well. In this case, instead of the structure in \eqref{eq:kalman} where the gain $\mf{L}[k]$ depends on $\hat{\mf{P}}[k]$, we use a gain $\mf{L}^{p_k}\in\{\mf{L}^{1},\dots,\mf{L}^D\}$ so that the artifact filter takes the form:
\begin{equation}
\label{eq:switched}
    \mf{\hat{x}}_s[k+1] = \mf{A}_d(\Delta^{p_k})\hat{\mf{x}}_s[k] + \mf{L}^{p_k}(\mf{z}[k] - \mf{C}\mf{\hat{x}}_s[k])
\end{equation}
with $\hat{\mf{x}}_s[0]=\mf{x}_0$. Hence, we require to design the gains $\mf{L}^{p_k}$ to make the filter asymptotically stable, making it feasible for the bound to exist. This cannot be done by designing each $\mf{L}^{p_k}$ by separate. The reason is that, the error $\tilde{\mf{x}}_s[k] = \mf{x}[k] - \mf{\hat{x}}_s[k]$ for \eqref{eq:switched} satisfies
\begin{equation}
\label{eq:switched2}
 \tilde{\mf{x}}_s[k+1] = \bm{\Lambda}^{p_k}  \tilde{\mf{x}}_s[k] + \mf{L}^{p_k}\mf{v}[k] + \mf{w}_d[k]
\end{equation}
with $ \bm{\Lambda}^{p_k} = \mf{A}_d(\Delta^{p_k})+\mf{L}^{p_k}\mf{C}$. Hence, \eqref{eq:switched2} is a switched system which switches between system matrices $\bm{\Lambda}^{p_k}$ with the schedule $p_k$ as switching signal. In the following auxiliary technical lemmas, we aim to ensure asymptotic stability of the filter, by designing $\mf{L}^{p_k}$ through the method of the common Lyapunov function \cite[Page 22]{liberzon2003}.

\label{app:lmi}
\begin{lemma}
\label{lem:cov_dyn}
Consider the filter \eqref{eq:switched} for some perception schedule $p$. Then, $\bm{\varrho}_s[k]:=\vect(\hat{\mf{P}}_s[k])$ with $\hat{\mf{P}}_s[k]:=\cov\{\mf{x}[k]-\mf{x}_s[k]\}$ satisfies:
\begin{equation}
\begin{aligned}
\label{eq:cov_dyn_s}
    \bm{\varrho}_s[k+1] &= (\bm{\Lambda}^{p_k}\otimes \bm{\Lambda}^{p_k})\bm{\varrho}_s[k] + \bm{\omega}[k], \bm{\varrho}_s[0] &= \vect(\mf{P}_0)
\end{aligned}
\end{equation}
where $\bm{\omega}[k]=\vect((\mf{L}^{p_k})\mf{R}^{p_k}(\mf{L}^{p_k})^\top+\mf{W}_d(\Delta^{p_k}))$.
\end{lemma}
\begin{pf}
First, compute $\hat{\mf{P}}_s[k+1]$ by applying $\cov(\bullet)$ to both sides of \eqref{eq:switched2} as:
\begin{equation*}
\begin{aligned}
    \hat{\mf{P}}_s[k+1]&=(\bm{\Lambda}^{p_k})\hat{\mf{P}}_s[k] (\bm{\Lambda}^{p_k})^\top\\&+ (\mf{L}^{p_k})\mf{R}^{p_k}(\mf{L}^{p_k})^\top+\mf{W}_d(\Delta^{p_k})
    \end{aligned}
\end{equation*}
Then, apply $\vect(\bullet)$ to both sides of the previous equation as well as the identity $\vect((\bm{\Lambda}^{p_k})\hat{\mf{P}}_s[k] (\bm{\Lambda}^{p_k})^\top) \equiv (\bm{\Lambda}^{p_k}\otimes \bm{\Lambda}^{p_k})\vect(\hat{\mf{P}}_s[k])$ to obtain \eqref{eq:cov_dyn_s}. Finally, $\hat{\mf{P}}[0] = \cov\{\mf{x}[0]-\mf{x}_0\} = \mf{P}_0$.
\end{pf}

\begin{lemma}
\label{lem:nonlinear_mi}
Consider $\bm{\Omega}, \gamma, \mf{Y}^i, i\in\{1,\dots,D\}$ satisfy \eqref{eq:lmi}. Henceforth, the following nonlinear matrix inequality is satisfied:
\begin{equation}
\label{eq:lmi2}
   \left( (\bm{\Lambda}^{p_k})^\top\bm{\Omega} (\bm{\Lambda}^{p_k})\otimes (\bm{\Lambda}^{p_k})^\top\bm{\Omega} (\bm{\Lambda}^{p_k})\right)\preceq \gamma^2 (\bm{\Omega}\otimes \bm{\Omega})
\end{equation}
with $\bm{\Lambda}^{i}=\mf{A}_d(\Delta^i)-\mf{L}^i\mf{C}$ and $\mf{L}^i=\bm{\Omega}^{-1}\mf{Y}^i, i\in\{1,\dots,D\}$.
\end{lemma}
\begin{pf}
Equivalence between \eqref{eq:lmi} and $(\bm{\Lambda}^{p_k})^\top\bm{\Omega} (\bm{\Lambda}^{p_k})\preceq \gamma \bm{\Omega}$ follows from the well known relationship of the Schur complement similarly as in \cite{lmi}. Now, apply Lemma \ref{lem:kron_def} with $\mf{M}_1=(\bm{\Lambda}^{p_k})^\top\bm{\Omega} (\bm{\Lambda}^{p_k})$ and $\mf{M}_2=\gamma \bm{\Omega}$ to obtain \eqref{eq:lmi2}.
\end{pf}

\begin{lemma}
\label{le:lyap}
Consider the assumptions in Lemmas \ref{lem:cov_dyn} and \ref{lem:nonlinear_mi}. Moreover, let $V(\bm{\varrho}_s[k]):=\sqrt{\bm{\varrho}_s[k]^\top(\bm{\Omega}\otimes \bm{\Omega})\bm{\varrho}_s[k]}$. Then, the following inequality is satisfied:
\begin{equation}
\label{eq:lyap}
    V(\bm{\varrho}_s[k+1]) \leq \gamma V(\bm{\varrho}_s[k])+\lambda_{\max}(\bm{\Omega})\overline{G}
\end{equation}
where $\lambda_{\max}(\bm{\Omega})$ is the maximum eigenvalue of $\bm{\Omega}$ and $$\overline{G}:=\max_{i\in\{1,\dots,D\}}\|(\mf{L}^{p_k})\mf{R}^{p_k}(\mf{L}^{p_k})^\top~+~\mf{W}_d(\Delta^{p_k})\|_F $$.
\end{lemma}
\begin{pf}
First, note that since $\bm{\Omega}$ is positive definite, henceforth $\|\bullet\|_{\bm{\Omega}\otimes \bm{\Omega}}:=\sqrt{(\bullet)^\top(\bm{\Omega}\otimes \bm{\Omega})(\bullet)}$ is a norm \cite[Page 321]{horn} and $V(\bm{\varrho}_s[k])\equiv\|\bm{\varrho}_s[k]\|_{\bm{\Omega}\otimes\bm{\Omega}}$. Now compute $V(\bm{\varrho}_s[k+1])$ from \eqref{eq:cov_dyn_s} as:
\begin{equation*}
    \begin{aligned}
    V(\bm{\varrho}_s[k+1])&= \|(\bm{\Lambda}^{p_k}\otimes \bm{\Lambda}^{p_k})\bm{\varrho}_s[k] + \bm{\omega}[k]\|_{\bm{\Omega}\otimes\bm{\Omega}} \\
    &\leq\|(\bm{\Lambda}^{p_k}\otimes \bm{\Lambda}^{p_k})\bm{\varrho}_s[k]\|_{\bm{\Omega}\otimes\bm{\Omega}} + \|\bm{\omega}[k]\|_{\bm{\Omega}\otimes\bm{\Omega}}
    \end{aligned}
\end{equation*}
using the triangle inequality \cite[Definition 5.1.1-(3)]{horn}. Furthermore, use Lemma \ref{lem:nonlinear_mi} to obtain:
\begin{equation}
\begin{aligned}
&\|(\bm{\Lambda}^{p_k}\otimes \bm{\Lambda}^{p_k})\bm{\varrho}_s[k]\|_{\bm{\Omega}\otimes\bm{\Omega}}\\& = \sqrt{\bm{\varrho}_s[k]^\top\left( (\bm{\Lambda}^{p_k})^\top\bm{\Omega} (\bm{\Lambda}^{p_k})\otimes (\bm{\Lambda}^{p_k})^\top\bm{\Omega} (\bm{\Lambda}^{p_k})\right)\bm{\varrho}_s[k]}\\
&\leq\sqrt{\gamma^2\bm{\varrho}_s[k]^\top(\bm{\Omega}\otimes \bm{\Omega})\bm{\varrho}_s[k]} \\
&= \gamma \|\bm{\varrho}_s[k]\|_{(\bm{\Omega}\otimes \bm{\Omega})}\equiv\gamma V(\bm{\varrho}_s[k])
\end{aligned}
\end{equation}
Moreover, note that $\bm{\omega}[k]^\top(\bm{\Omega}\otimes \bm{\Omega})\bm{\omega}[k]\leq \lambda_{\max}(\bm{\Omega}\otimes \bm{\Omega})\|\bm{\omega}[k]\|^2$ by the Rayleigh inequality \cite[Theorem 4.2.2]{horn}. Furthermore, note that $$\|\bm{\omega}[k]\|=\|(\mf{L}^{p_k})\mf{R}^{p_k}(\mf{L}^{p_k})^\top+\mf{W}_d(\Delta^{p_k})\|_F.$$ Thus, $\|\bm{\omega}[k]\|_{\bm{\Omega}\otimes\bm{\Omega}}\leq \lambda_{\max}(\bm{\Omega})\overline{G}$. Then, \eqref{eq:lyap} is the combination of the previous results.
\end{pf}
We are now ready to show the rest of Proposition \ref{eq:finite_time}. First, consider a scalar signal $v[k]\in\mathbb{R}$ satisfying:
$$
v[k+1] = \gamma v[k] + \lambda_{\max}(\bm{\Omega})\overline{G}, 
$$
with $\lambda_{\max}(\bm{\Omega})\overline{G}$ as in Lemma \ref{le:lyap}, $\gamma\in(0,1)$ and $v[0]=\sqrt{\vect(\mf{P}_0)^\top(\bm{\Omega}\otimes\bm{\Omega})\vect(\mf{P}_0})$. It can be verified that the solution to the previous linear difference equation satisfies:
$$
\begin{aligned}
v[k] &= \gamma^k v[0] + \lambda_{\max}(\bm{\Omega})\overline{G}\left(\frac{1-\gamma^k}{1-\gamma}\right)\\
&\leq v[0]+\frac{\lambda_{\max}(\bm{\Omega})\overline{G}}{1-\gamma}, \forall k\geq 0
\end{aligned}
$$
In addition, note that 
$$
v[0]\leq \sqrt{\lambda_{\max}(\bm{\Omega}\otimes\bm{\Omega})\vect(\mf{P}_0)^\top\vect(\mf{P}_0)} = \lambda_{\max}(\bm{\Omega})\|\mf{P}_0\|_F
$$
by means of the Rayleigh inequality \cite[Theorem 4.2.2]{horn} and the identity $\|\vect(\mf{P}_0)\|\equiv\|\mf{P}_0\|_F$. Combine the previous results to conclude that $v[k]\leq B_s', \forall k\geq 0$ with $B_s'=\lambda_{\max}(\bm{\Omega})\left(\|\mf{P}_0\|_F+\frac{\overline{G}}{1-\gamma}\right)$. Recall that $V(\bm{\varrho}_s[k])$ complies \eqref{eq:lyap} by Lemma \ref{le:lyap}. Use the comparison lemma in \cite[Lemma 13]{comparison} together with $V(\bm{\varrho}_s[0])=v[0]$ to conclude that $V(\bm{\varrho}_s[k])\leq v[k], \forall k\geq 0$ which leads directly $V(\bm{\varrho}_s[k])\leq B_s'$. Now, Rayleigh inequality is used again to conclude that
$
\lambda_{\min}(\bm{\Omega})\|\bm{\varrho}_s[k]\|\leq V(\bm{\varrho}_s[k]) \leq B_s'
$
equivalently $\|\bm{\varrho}_s[k]\|=\|\hat{\mf{P}}_s[k]\|_F\leq {B_s'}/{\lambda_{\min}(\bm{\Omega})}\equiv B_s/\sqrt{n_{\mf{x}}}$ with $B_s$ defined in \eqref{eq:switched_bound}. Use the same arguments as in the the proof of Lemma \ref{th:astorm_kalman} to conclude that $\hat{\mf{P}}[k]\preceq \hat{\mf{P}}_s[k]$ where $\hat{\mf{P}}[k]$ is obtained from PLATE in \eqref{eq:kalman}. Hence, use the same arguments as in the proof of Lemma \ref{le:bounds}-b) in \ref{app:aux} to obtain $\|\hat{\mf{P}}[k]\|_F\leq \sqrt{n_{\mf{x}}}\|\hat{\mf{P}}_s[k]\|_F\leq B_s$.

\subsection{Proof of Proposition \ref{prop:optimal_approx}}
\label{ap:optimal_approx}

Similar to the proof of Proposition \ref{prop:worst_case}, optimality of the solution of Algorithm \ref{algo:approx_2} comes from the dynamic programming principle \cite[Page 23]{bertsekas2000}. As for the complexity, note that combining lines 6, 7 and 9 of Algorithm \ref{algo:approx} results in $\alpha_{\max}Q(\delta)D$ iterations needed to compute $\mf{M}_P,\mf{M}_J,M_\alpha$. In addition, the number of steps in lines 2 and 5 in Algorithm \ref{algo:approx_2} are $Q(\delta)$ and $\alpha_{\max}$ respectively. Hence, the asymptotic complexity of Algorithm \ref{algo:approx_2} is only given by the term $\alpha_{\max}Q(\delta)D$.

\subsection{Proof of Theorem \ref{th:closedness}}
\label{app:cost}
In this section, we denote with $\mf{F}^{p_k}$ the function that computes $\hat{\mf{P}}[k+1]=\mf\mf{F}^{p_k}(\hat{\mf{P}}[k])$ according to \eqref{eq:kalman}. Moreover, in order to show Theorem \ref{th:closedness}, we provide an auxiliary lemma.
\begin{lemma}
\label{eq:continous_dependence}
Let a given $K\in\mathbb{N}$ and consider the systems:
$$
\hat{\mf{P}}[k+1]=\mf\mf{F}^{p_k}(\hat{\mf{P}}[k]), \ \ \hat{\mf{P}}[0] = \mf{P}_0
$$
$$
\hat{\mf{P}}'[k+1]=\mf\mf{F}^{p_k}(\hat{\mf{P}}'[k]) + \bm{\Pi}_{\mathcal{Q}}[k], \ \ \hat{\mf{P}}'[0] = \mf{P}_0'
$$
with $\|\hat{\mf{P}}[0]-\hat{\mf{P}}'[0]\|_F\leq \delta$ and $\|\bm{\Pi}_{\mathcal{Q}}[k]\|_F\leq \delta, \forall k\in\{1,\dots,K\}$. Therefore, for any $\varepsilon'>0$ there exists $\delta>0$ such that $\|\hat{\mf{P}}[k]-\hat{\mf{P}}'[k]\|_F\leq \varepsilon', \forall k\in\{1,\dots,K\}$. 
\end{lemma}
\begin{pf}
First, note that for any $\varepsilon_0>0$, there exists $\delta>0$ such that $\|\hat{\mf{P}}[1]-\hat{\mf{P}}'[1]\|_F=\|\mf{F}^{p_0}(\hat{\mf{P}}[0])-\mf{F}^{p_0}(\hat{\mf{P}}'[0])-\bm{\Pi}_\mathcal{Q}[0]\|_F\leq \|\mf{F}^{p_0}(\hat{\mf{P}}[0])-\mf{F}^{p_0}(\hat{\mf{P}}'[0])\|_F+\delta\leq \varepsilon_0$ due to continuity of $\mf{F}^{p_0}(\bullet)$. The same reasoning applies for the remaining $K-1$ steps, making $\hat{\mf{P}}[k]$ arbitrarily close to $\hat{\mf{P}}'[k]$ by choosing $\delta>0$ sufficiently small.
\end{pf}
Using the previous result, the proof of Theorem \ref{th:closedness} follows. First note that the initial condition $\mf{P}_0$ for the original problem and the quantized version $\hat{\mf{P}}_{q_0} = \mathcal{Q}(\mf{P}_0)$ comply $\| \mf{P}_0 - \hat{\mf{P}}_{q_0} \|_F\leq \delta$.  Note that if $\hat{\mf{P}}[k]\in\{\hat{\mf{P}}_1,\dots,\hat{\mf{P}}_{Q(\delta)}\}$, then $\left\|\mf{F}^{p_k}(\hat{\mf{P}}[k]) - \mathcal{Q}\left(\mf{F}^{p_k}(\hat{\mf{P}}[k])\right)\right\|_F\leq \delta$. Thus, in the quantized setting, $\hat{\mf{P}}'[k]$ evolves according to $\hat{\mf{P}}'[k+1] = \mf\mf{F}^{p_k}(\hat{\mf{P}}'[k]) + \bm{\Pi}_{\mathcal{Q}}[k]$ where $\bm{\Pi}_{\mathcal{Q}}[k]$ is the quantization noise complying $\|\bm{\Pi}_{\mathcal{Q}}[k]\|_F\leq \delta$ and initial condition $\hat{\mf{P}}'[0]=\hat{\mf{P}}_{q_0}$. Thus, Lemma \ref{eq:continous_dependence} implies that for any $\varepsilon'>0$ there is sufficiently small $\delta>0$ such that true optimal trajectory for the covariance $\hat{\mf{P}}[k]$ and the quantized one $\hat{\mf{P}}'[k]$ comply $\|\hat{\mf{P}}[k]-\hat{\mf{P}}'[k]\|_F\leq \varepsilon'$ equivalently $\hat{\mf{P}}[k] - \hat{\mf{P}}'[k] = \tilde{\mf{P}}[k]$ for some $\tilde{\mf{P}}[k]$ with $\|\tilde{\mf{P}}[k]\|_F\leq \varepsilon'$. Now, with $\tau_{k+1}^+:=\min(\tau_{k+1},T_f)$:
$$
\begin{aligned}
&|\mathcal{J}-\mathcal{J}_{\mathcal{Q}}(\delta)|\\
&\leq \frac{1}{T_f}\sum_{k=0}^{\len(p)}\int_{\tau_k}^{\tau_{k+1}^+}\left|\tr(\mf{A}_d(t-\tau_k)\tilde{\mf{P}}[k]\mf{A}_d(t-\tau_k)^\top)\right|\text{d}t\\
&\leq \frac{1}{T_f}\sum_{k=0}^{\len(p)}\left(\tau_{k+1}^+-\tau_k\right) \sup_{\tau_{k}\leq \tau\leq \tau_{k+1}}\left|\tr(\mf{A}_d(\tau)\tilde{\mf{P}}[k]\mf{A}_d(\tau)^\top)\right| \\
&\leq \max_{0\leq k\leq \len(p)}\sup_{\tau_{k}\leq \tau\leq \tau_{k+1}}\left|\tr(\mf{A}_d(\tau)\tilde{\mf{P}}[k]\mf{A}_d(\tau)^\top)\right|
\end{aligned}
$$
However, since $\|\tilde{\mf{P}}[k]\|_F\leq \varepsilon'$, one can choose $\delta>0$ sufficiently small to make $\varepsilon'>0$ and as a consequence $\left|\tr(\mf{A}_d(\tau)\tilde{\mf{P}}[k]\mf{A}_d(\tau)^\top)\right|$ to be arbitrarily small for any $\tau\in[\tau_k,\tau_{k+1}], k\in\{1,\dots,\len(p)\}$. Thus, $|\mathcal{J} - \mathcal{J}_{\mathcal{Q}}(\delta)|\leq \varepsilon$ for some $\delta>0$.

\bibliographystyle{elsarticle-num}

\end{document}